\documentclass[twocolumn]{aastex631}
\usepackage{kotex}
\usepackage{soul}
\usepackage{multirow}

\received{2022 January 10}
\revised{2022 July 11}
\accepted{2022 July 12}
\published{2022 August 25}

\shorttitle{Nonlinear Color--Metallicity Relations of Globular Clusters. XI.}
\shortauthors{Kim et al.}
\graphicspath{{./}{figures/}}

\begin{document}

\title{Nonlinear Color$-$Metallicity Relations of Globular Clusters. XI. Nonlinearity Effect Revealed by NGC 5128 (Centaurus A) and NGC 4594 (Sombrero) Galaxies}

\correspondingauthor{Hak-Sub Kim}
\email{agapiel96@gmail.com}

\author[0000-0001-7033-4522]{Hak-Sub Kim}
\altaffiliation{Both authors have contributed equally to this paper.}
\affiliation{Department of Physics and Astronomy, Sejong University, Seoul 05006, Republic of Korea}

\author[0000-0002-1842-4325]{Suk-Jin Yoon}
\altaffiliation{Both authors have contributed equally to this paper.}
\affiliation{Department of Astronomy \& Center for Galaxy Evolution Research, Yonsei University, Seoul 03722, Republic of Korea}

\author[0000-0002-7957-3877]{Sang-Yoon Lee}
\affiliation{Department of Astronomy \& Center for Galaxy Evolution Research, Yonsei University, Seoul 03722, Republic of Korea}

\author{Sang-Il Han}
\affiliation{Department of Science Education, Ewha Womans University, Seoul 03760, Republic of Korea}

\begin{abstract}

Metallicity distributions (MDs) of globular clusters (GCs) provide crucial clues for the assembly and star formation history of their host galaxies. 
GC colors, when GCs are old, have been used as a proxy of GC metallicities. 
Bimodal GC color distributions (CDs) observed in most early-type galaxies have been interpreted as bimodal MDs for decades, suggesting the presence of merely two GC subpopulations within single galaxies. 
However, the conventional view has been challenged by a new theory that nonlinear metallicity-to-color conversion can cause bimodal CDs from unimodal MDs. 
The unimodal MDs seem natural given that MDs involved many thousand protogalaxies. 
The new theory has been tested and corroborated by various observational and theoretical studies.

Here we examine the nonlinear nature of GC color$-$metallicity relations (CMRs) using photometric and spectroscopic GC data of NGC 5128 (Centaurus A) and NGC 4594 (Sombrero), in comparison with stellar population simulations.
We find that, with a slight offset in color, the overall shapes of observed and modeled CMRs agree well for all available colors.
Diverse color-depending morphologies of GC CDs of the two galaxies are well reproduced based on their observed spectroscopic MDs via our CMR models. 
The results corroborate the nonlinear CMR interpretation of the GC color bimodality, shedding further light on theories of galaxy formation.

\end{abstract}

\section{Introduction} \label{sec:intro}

Globular clusters (GCs) are among the oldest stellar systems in the Universe.
They are thought to have formed during the major star formation events in galactic history and survived relatively intact throughout the galactic evolution.
Since they retain the chemical properties of the surrounding environment at the formation epoch, their metallicity distribution (MD) provides a vital clue to the formation and evolution of galaxies \citep{Bastian18}.
Due to the limitations of observational instruments, GC MDs are often inferred from photometric data rather than obtained directly through spectroscopy.
GC colors have been used as proxies for GC metallicity under the assumption that, when GCs are old, colors correlate linearly with metallicity.

One of the most important features observed in GC color distribution (CD) is bimodality,
in that the CD of GCs belonging to one galaxy exhibits two peaks \citep{Brodie06}.
Since this phenomenon does not appear only in certain galaxies, 
but in most large galaxies, proper interpretation of this phenomenon
is crucial to the study of galaxy formation and evolution.
With empirical linear GC color--metallicity relations (CMRs), 
bimodal CDs have generally been translated to bimodal MDs, 
implying the presence of two populations of GCs within individual galaxies
\citep[e.g.,][]{Park13, Lee16, Forbes18, Bardy19}. 
However, \citet[hereafter Paper I]{Yoon06} give a plausible alternative explanation 
for the origin of GC color bimodality, in which the nonlinear nature of GC CMRs 
can lead to the bimodal CDs even from a single-peaked MD.
In this theory, the underlying GC MDs could be quite different
from that of GC CDs characterized by bimodality.
Although many studies have attempted to examine the new explanation
by various means \citep[e.g.,][]{Cantiello07, Kundu07, Strader07, Peng08, Spitler08, Blakes10, Schuberth10, Blakes12, Brodie12, Chies12, Park12, Pota13, Park13, Vanderbeke14, Cantiello14, Richtler15, Salinas15, Cho16, Vill19, Fahrion20},
this issue is still under debate. 
Elucidating what the nonlinearity is like in GC CMR and how it affects the color-to-metallicity transformation is at the heart of this debate.

Since Paper I, a series of papers \citep[hereafter Papers II $-$ X]{Yoon11a, Yoon11b, Yoon13, Kim13, Chung16, Kim17, Lee19, Lee20, Kim21} have been scrutinized the nonlinear CMR theory, both observationally and theoretically. 
Papers II and IV reported strong color-dependent variations in the CD morphology of M87 and of M84 GCs, respectively, which were incomprehensible with linear, straight CMRs but were well reproduced by the nonlinear CMRs. 
Paper III proposed a highly viable solution based on the CMR nonlinearity for the perplexing conundrum of the MD discrepancy between GCs and field halo stars in given parent galaxies. 
Papers V and VII paid attention to bimodal distributions of absorption-line indices such as Balmer and Mg$b$ lines for GCs in highly studied M31 and NGC 5128, respectively.
The bimodality is readily explained by nonlinear index$-$metallicity relations, analogous to the nonlinear CMR projection. 
Paper VI approached theoretically to the MDs derived using the Ca Triplet index, a widely used metallicity indicator.
They showed that the Ca Triplet$-$color$-$metallicity relation is highly nonlinear, resulting in a sharp difference between color and Ca Triplet distributions of old GCs.
Paper VIII proposed a simple, coherent explanation for the observed diversity in the GC CDs of early-type galaxies in the Virgo and Fornax galaxy clusters, where the diversity is due mainly to changes in the mean metallicity of GC systems.
Paper IX explained the difference in the radial number density profile between blue and red GCs by combining the radial metallicity gradient of the GC system and the nonlinear nature of metallicity-to-color transformation.
Paper X observed GCs in M87 with the Subaru/FOCAS multiobject spectroscopy mode.
Their high-quality, homogeneous dataset corroborated the theoretical prediction of inflected CMRs.

As part of the series papers, we investigate the presence and degree of nonlinearity along GC CMRs for various colors by combining multiband photometric data with spectroscopic data of the two nearby massive galaxies, NGC 5128 (Centaurus A) and NGC 4594 (Sombrero).
The photometry of NGC 4594 GCs is from our observations and the other data are taken from the literature.
The NGC 5128, the central giant elliptical galaxy in the Centaurus group of galaxies, 
is one of the best targets to study the nature of GC CMRs thanks to its proximity ($\sim$4 Mpc),
the richness of GCs \citep[$>3000$;][]{Taylor17}, and relatively low differential reddening of E($B-V$)=0.07 \citep{Rejk02}.
The NGC 4594 is a lenticular galaxy with an unusually large bulge and a prominent dust lane 
in the nearly edge-on disk. Since it is also nearby ($\sim$9.55 Mpc) and has a relatively large number of GCs \citep[$\sim$1150;][]{Lars01},
this galaxy's GC system is well suited for GC CMR studies.

\section{Observational Data and Theoretical Model} \label{sec:data}
\subsection{NGC 5128 (Centaurus A) Globular Clusters}

The photometry for GCs in NGC 5128 was obtained from the compilation of \citet{Wood07}, which combined the $UBVRI$ photometry by \citet{Peng04} and the $CMT1$ photometry by \citet{Har04}.
We used the $UBVRI$ data because CMRs of $CMT1$ have too large scatter to examine the CMR nonlinearity. 
To compare the photometry with theoretical predictions, we applied a correction for Galactic extinction based on the COBE/DIRBE dust maps \citep{Schlegel98} with the correction terms by \citet{Sch11} and the extinction law by \citet{Cardelli89}.

\begin{figure}
\epsscale{1.2}
\plotone{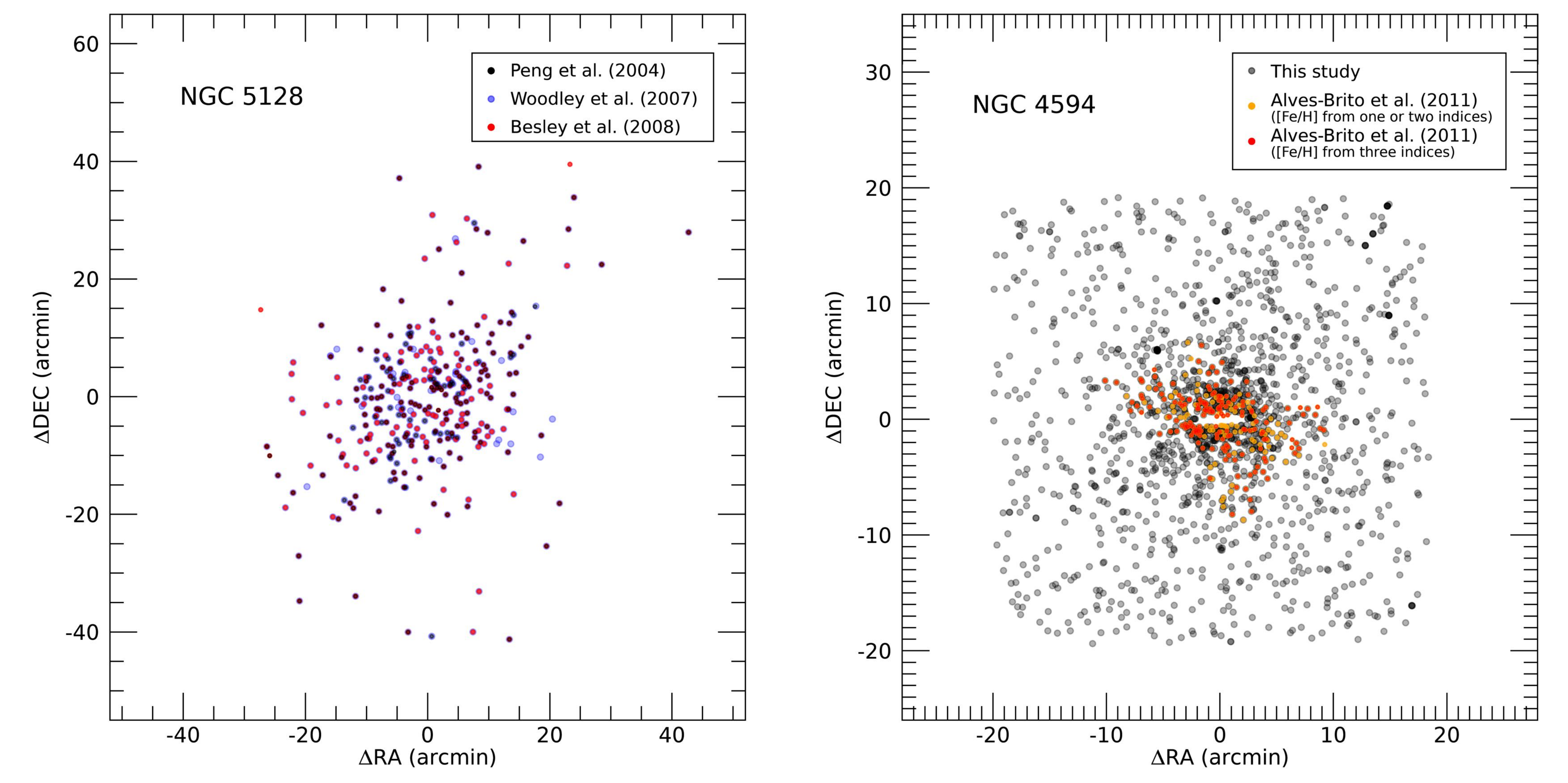}
\caption{
Spatial distribution of GC data set used in this study for NGC 5128 (left) and NGC 4594 (right). For NGC 5128, $UBVRI$ photometry is taken from \citet[black dots]{Peng04} and \citet[blue dots]{Wood07}, and spectroscopic metallicity is taken from \citet[red dots]{Beas08}. A total of 234 GCs are crossmatched between the photometry and the spectroscopy. For NGC 4594, $UBVI$ photometry is taken from our observations. Gray dots are GC candidates selected using two-color diagrams (see Section~\ref{sec:GCselect}). Spectroscopic metallicity is taken from \citet{Alv11}, who provide spectroscopic metallicity derived from three metallicity-sensitive indices (CH, Mg$b$, \& Fe5270). We only use the metallicities derived from all three indices (red dots), except for metallicities from one or two indices (orange dots). A total of 108 GCs are crossmatched between the photometry and the spectroscopy.
\label{fig:spatial}}
\end{figure}

The spectroscopic data were taken from \citet{Beas08}. 
They obtained integrated light spectra for 254 GCs in NGC 5128 using the 2-degree Field instrument on the Anglo-Australian Telescope and provided 21 Lick indices for 146 GCs and empirical spectroscopic metallicities for 243 GCs. 
The left panel of Figure~\ref{fig:spatial} shows the spatial distribution of photometric data of \citet[black dots]{Peng04} and~\citet[blue dots]{Wood07} and spectroscopic data of~\citet[red dots]{Beas08}.

We crossmatched these data using a matching radius of 1\arcsec,
resulting in a total number of 234 GCs with both the $UBVRI$ photometry and spectroscopic metallicity. 
The empirical metallicity ([M/H]) provided by \citet{Beas08} was transformed to [Fe/H] by subtracting 0.217 \citep{Kim02}.
Table~\ref{tab:n5128gc} gives astrometry, spectroscopic metallicity and $UBVRI$ photometry of crossmatched GC sample in NGC 5128.

\subsection{NGC 4594 (Sombrero) Globular Clusters}

We carried out $UBVI$ photometry for the GC system in NGC 4594 on the nights of 2008 May 4 and 5 (UT) with the Mosaic II CCD imager mounted on the prime focus of the 4 m Blanco telescope at Cerro Tololo Inter-American Observatory (CTIO). 
The observing conditions were clear and photometric with an average seeing of around 1$\arcsec$.
The Mosaic II consists of eight 2048$\times$4096 pixel CCDs with a pixel scale of 0\farcs27 providing a total field of view of 36$\arcmin\times$36$\arcmin$. 
The total exposure times in $U$, $B$, $V$, and $I$ were 18000, 4500, 3000, and 2000 seconds, respectively. 
We split our observations in each band into five dithered exposures (two sets for $U$) to fill in the gaps between the individual CCD chips in the mosaic, and to minimize the impact on our photometry from CCD blemishes such as hot pixels and bad columns.

All images were preprocessed using the MSCRED package \citep{Valdes98} in IRAF\footnote{IRAF is distributed by the National Optical Astronomy Observatory, which is operated by the Association of Universities for Research in Astronomy, Inc., under a cooperative agreement with the National Science Foundation.}.
We used the CCDPROC task to correct for cross-talk between the CCD chips, trim the overscan, correct for the bias level, and apply the flat-field correction.
The MSCPIXAREA task is used to correct for the variations in pixel scales across the frame produced by geometric distortion in the Mosaic II imager. 
Each mosaic image was then split into eight individual images using the MSCSPLIT task.
We combined the images using the MONTAGE2 in the ALLFRAME package \citep{Stet94} to produce a clean high signal-to-noise coadded image in each filter. 
Sources were detected from the combined images using the DAOPHOT II/ALLSTAR \citep{Stet87}, producing a master catalog of the sources. 
The master catalog was then input to the ALLFRAME along with all of the images and their point-spread function (PSF) models for final PSF photometry.
Aperture corrections were computed through the DAOGROW \citep{Stet90} using bright and isolated point sources, and we applied them to our photometry above.
Finally, an error-weighted mean instrumental magnitude for each object was obtained using the DAOMATCH/DAOMASTER \citep{Stet93}.

The photometric calibration was achieved using standard stars in the fields of \citet{Lan92} and \citet{Stet00}. 
We corrected for the foreground Galactic extinction using the reddening maps by \citet{Schlegel98} with the correction terms by \citet{Sch11} and the extinction law by \citet{Cardelli89}. 
Astrometric solutions were calculated using the USNO-B 1.0 catalog stars \citep{Monet03}, resulting in the typical r.m.s. level of $\sim$\,0$\farcs$2.

The spectroscopic data were taken from \citet{Alv11}, who measured three metallicity-sensitive indices (CH, Mg$b$, \& Fe5270) for 247 GC candidates and transformed the measurements into metallicities using the \citet{Brodie90} method. 
They provided error-weighted mean metallicities for 135 GC candidates for which all three indices are measured. From the sample, we selected GCs using radial velocity and projected galactocentric distance (see Figure 3 in \citealt{Alv11}) and used their mean metallicities for our analysis. 
The right panel of Figure~\ref{fig:spatial} shows the spatial distribution of photometric GC candidates (see Section~\ref{sec:GCselect}, gray dots) and spectroscopic data (yellow and red dots).

We crossmatched our $UBVI$ catalog with the spectroscopic data using a matching radius of 1\arcsec, resulting in a total number of 122 GCs with both $UBVI$ photometry and spectroscopic metallicity. Table~\ref{tab:n4594gc} gives GC ID, coordinate, spectroscopic metallicity, and $UBVI$ photometry of crossmatched GC sample in NGC 4594.

\subsection{Theoretical Evolutionary Population Synthesis Model}
The theoretical models used in this study are constructed using the Yonsei Evolutionary Population Synthesis (YEPS) code \citep{Chung13b}. 
The YEPS model comprehensively deals with the effect of horizontal-branch stars of simple stellar populations, which are the main source causing the nonlinearity of GC CMRs. 
GC colors, when GCs are old, have been used as a proxy of GC metallicities. 
We use the YEPS model of the best-matching age (12.5 Gyr) for both NGC 5128 and NGC 4594.
We adopt the enhanced $\alpha$-elements ([$\alpha$/Fe] = 0.3) and solar-scaled abundance ratios for other elements. Readers are referred to \citet{Chung13b, Chung17, Chung20} for details of the ingredients and input parameters of the YEPS model.
Table~\ref{tab:yeps_12.5} provides the YEPS predictions of metallicity and {\it UBVRI} colors for the 12.5 Gyr simple stellar population, which is used in our present analysis.

\subsection{Selection of Old Globular Clusters}
\label{sec:GCselect}

In order to properly examine the nonlinearity of CMRs for old GCs, possible young and intermediate-age GCs should be excluded from the sample.
Both NGC 5128 and NGC 4594 have strong dust lanes across the central regions, hinting at recent wet merger events and subsequent star formation.
The cosmological hydrodynamical simulations by \citet{Du21} suggested that Sombrero-like galaxies (halo-dominated galaxies with huge bulges) have experienced residual star formation after major merger events during the formation history.
It is thus naturally expected that young/intermediate-age GCs are intermingled with the dominant old GCs of the galaxies.

The GC age from integrated light is usually estimated from the comparison of age-sensitive indices such as Balmer lines and metallicity-sensitive indices in spectroscopy \citep{Puzia05, Puzia06, Worthey94}, or from two-color diagrams of the combination of age- and metallicity-sensitive colors in photometry \citep{Puzia02, Hempel03, Yi04, deGrijs05}. 
We chose the photometric method to select old GCs from the crossmatched data to secure as a large number of GCs as possible for the CMR nonlinearity test.

\begin{figure*}
\epsscale{1.2}
\plotone{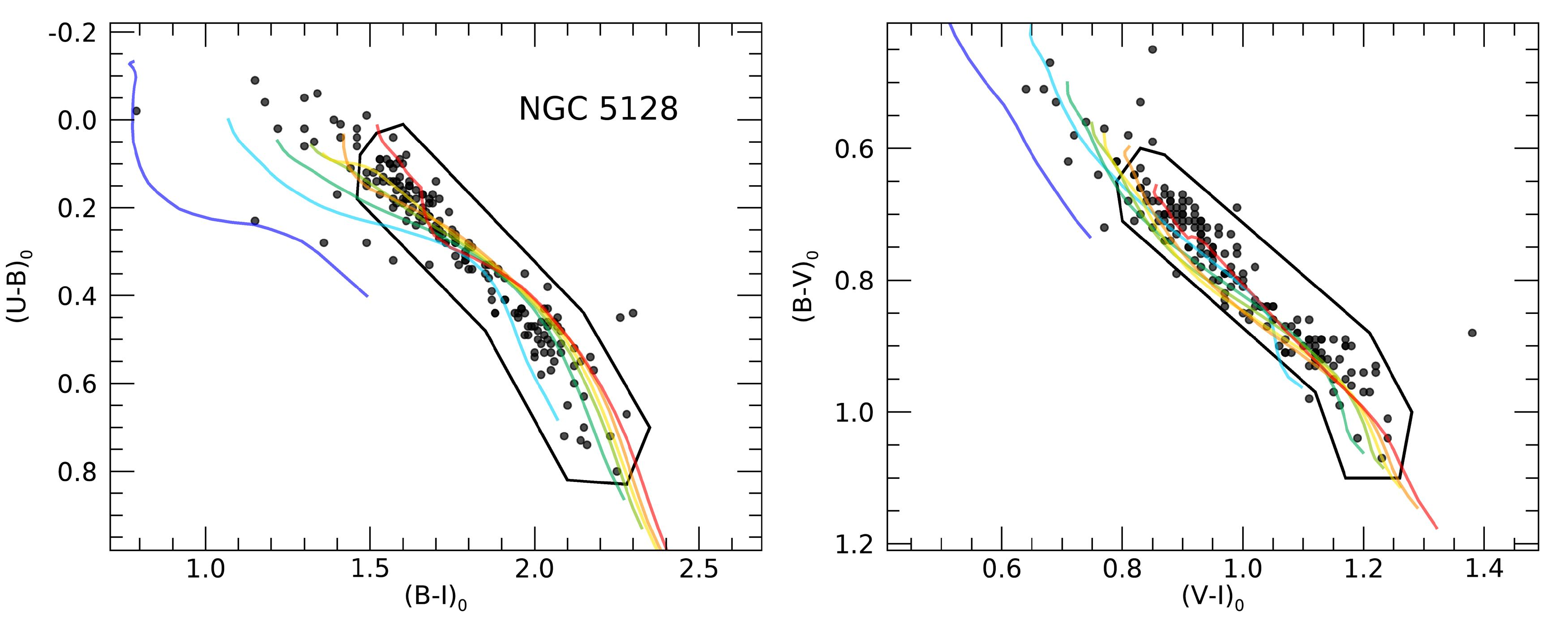}
\caption{
Two-color diagrams of crossmatched GC sample in NGC 5128. The YEPS model predictions for GCs with ages from 1 Gyr (blue) to 13 Gyr (red) are overlaid in 2 Gyr intervals. The model loci are slightly shifted to the direction of observations (see section~\ref{sec:CMRs}). We select old GCs using the polygons marked by thick black lines. 
Note that young GCs with intermediate and high metallicities are indistinguishable from old GCs given the observational uncertainty and are partially removed, contaminating the sample to some extent.
\label{fig:2col_5128}}
\end{figure*}

\begin{figure*}
\epsscale{1.2}
\plotone{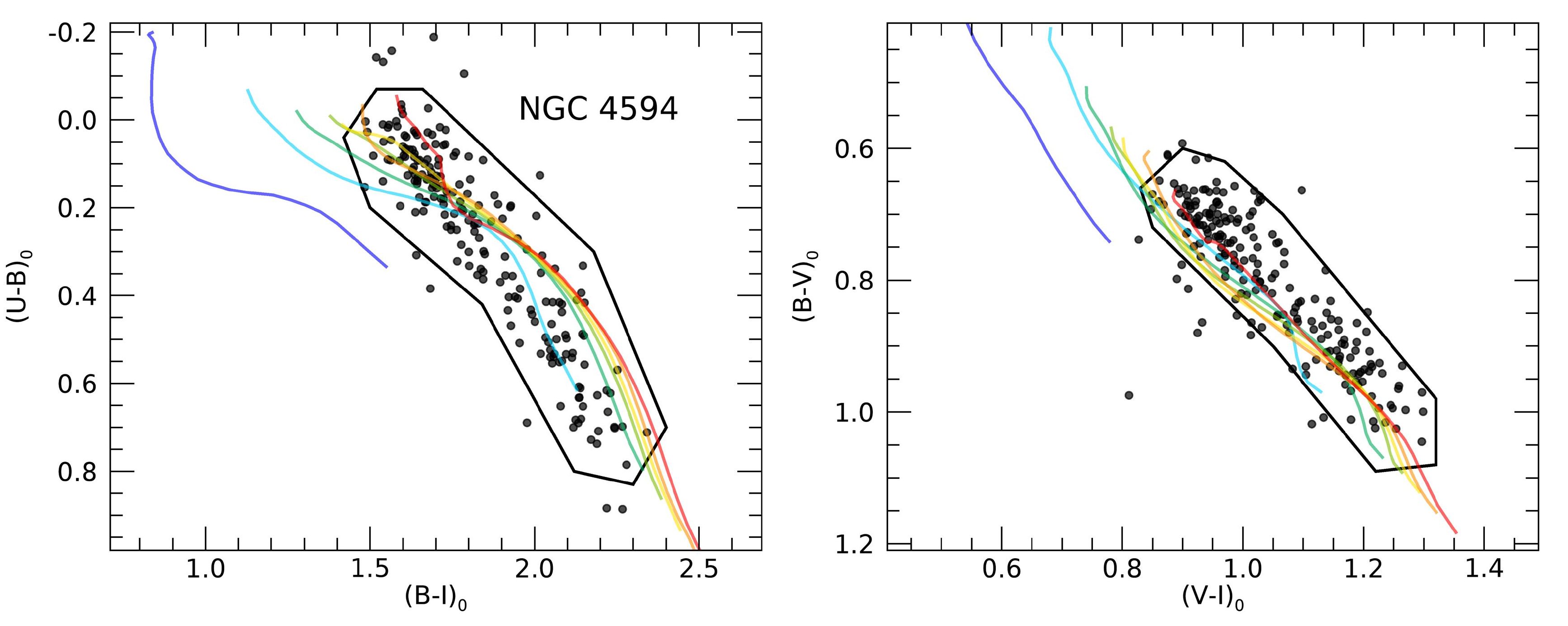}
\caption{
Same as Figure~\ref{fig:2col_5128}, but for crossmatched GC sample in NGC 4594.
\label{fig:2col_4594}}
\end{figure*}

Figures~\ref{fig:2col_5128} and~\ref{fig:2col_4594} show the two-color diagrams of GCs in NGC 5128 and in NGC 4594, respectively.
The selection region is denoted by the polygon drawing by thick black lines in each panel, and the YEPS models for ages from 1 Gyr (blue) to 13 Gyr (red) at 2 Gyr intervals are overplotted for reference.
The model loci are slightly shifted to the direction of observations using the measured offsets, which will be discussed in the following section.
The use of $U$- and $B$-band data enables us to discern young/intermediate GCs from old GCs, especially for metal-poor GCs, since young/intermediate GCs are bluer in $U-B$ and $B-V$ colors. 
We determine the selection regions to exclude these blue GCs in $U-B$ and $B-V$ colors and the outliers laid away from the GC clustering in the two-color diagrams.
However, we note that, as young GCs with intermediate and high metallicities are indistinguishable from old GCs given the observational uncertainties, they are eliminated only partially, contaminating the sample to some extent.

\section{Color--Metallicity Relations} \label{sec:CMRs}

It has been known that there are some offsets between observations and models in the color--magnitude planes \citep{Van03, Van10, Worthey11, Ang15} and in the integrated colors \citep{Barmby00, Lee02, Conroy09}.
The discrepancy stems mainly from the incompleteness of models, such as imperfect treatments of binaries, blue stragglers, horizontal-branch stars, and stars in the late evolutionary stages \citep{Chung13a, Chung17, Jang21}.
To compensate for the defect, we shift the model CMRs in the direction of colors to match the observed CMRs, 
allowing a more direct comparison of the model predictions with observations.
The amount of shifts is determined using the total least squares regression method with sigma clipping. 
After transforming colors and metallicities into a normalized format, we calculate the inversely error-weighted total least squares between observations and models by moving the model loci by the interval of 0.01 mag in the color direction.
We then find the amount of shifts, which minimizes the sum of the orthogonal residuals.

\begin{figure}
\epsscale{1.0}
\plotone{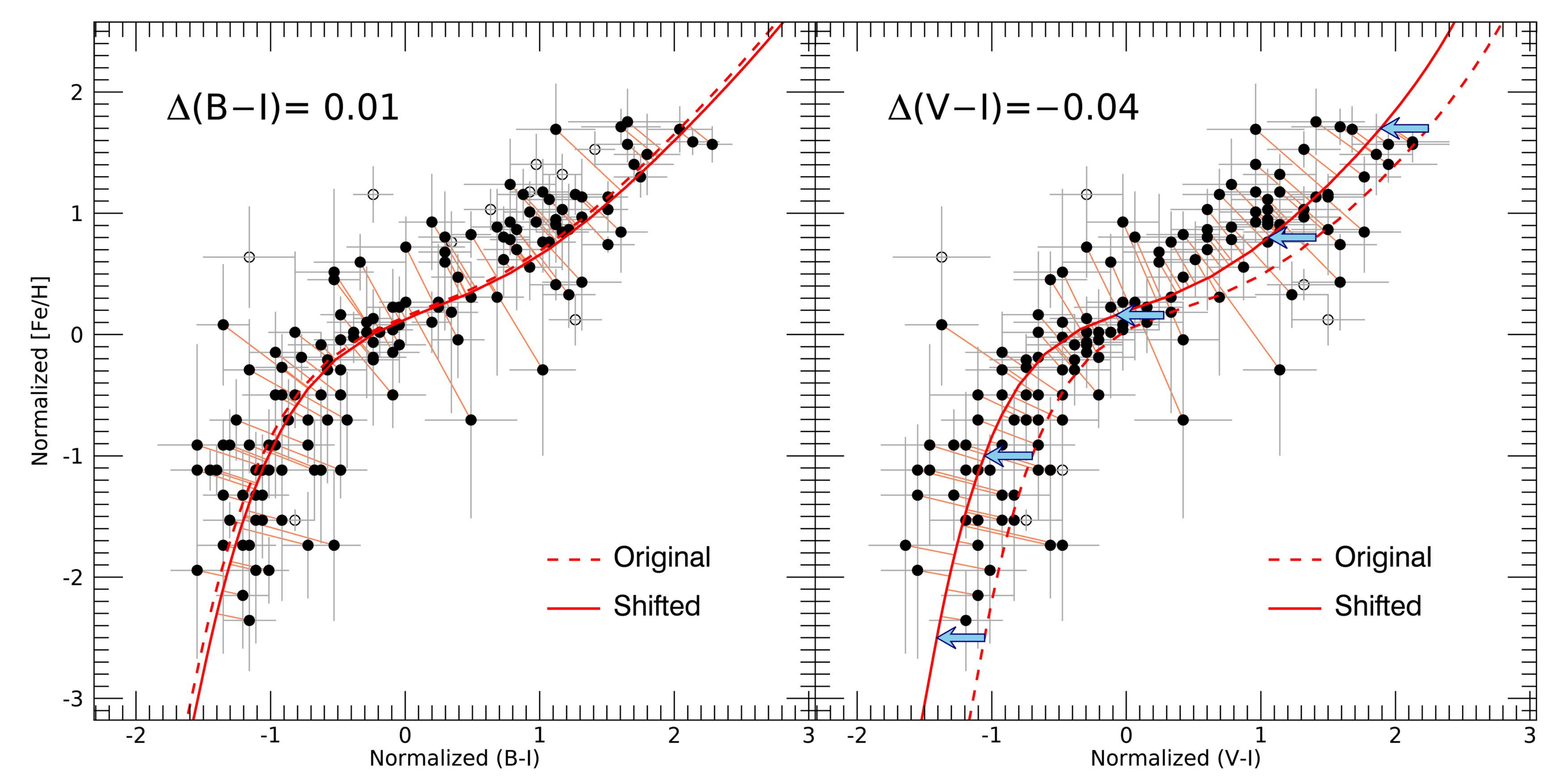}
\caption{
Determination of offsets between observations and model predictions for B-I GC CMR (left) and V-I GC CMR (right). The offset value denoted on the top-left corner of each panel is estimated by the weighted total least squares method on the normalized plane of colors and metallicities. The orange lines indicate orthogonal distances from the data points to the model CMR. The black circles are the final sample used in the total least squares method and the open circles are outliers removed by 5-sigma clipping. The dotted line represents the original model prediction of the GC CMR and the solid line shows the shifted model locus after applying the offset value, as shown by cyan horizontal arrows in the right panel.}
\label{fig:normal}
\end{figure}

Figure~\ref{fig:normal} shows an example of the process of determining the offset value between observations and models.
The dotted line represents the original model prediction of the GC CMR and the red solid line shows the shifted model locus after applying the offset value denoted in the top-left corner of the figure. 
The orange lines show the orthogonal distances from the data points to the model CMR.
The black circles are the final sample used in the total least squares method and the open circles are outliers removed by 5$\sigma$ clipping.

\begin{figure}
\epsscale{1.2}
\plotone{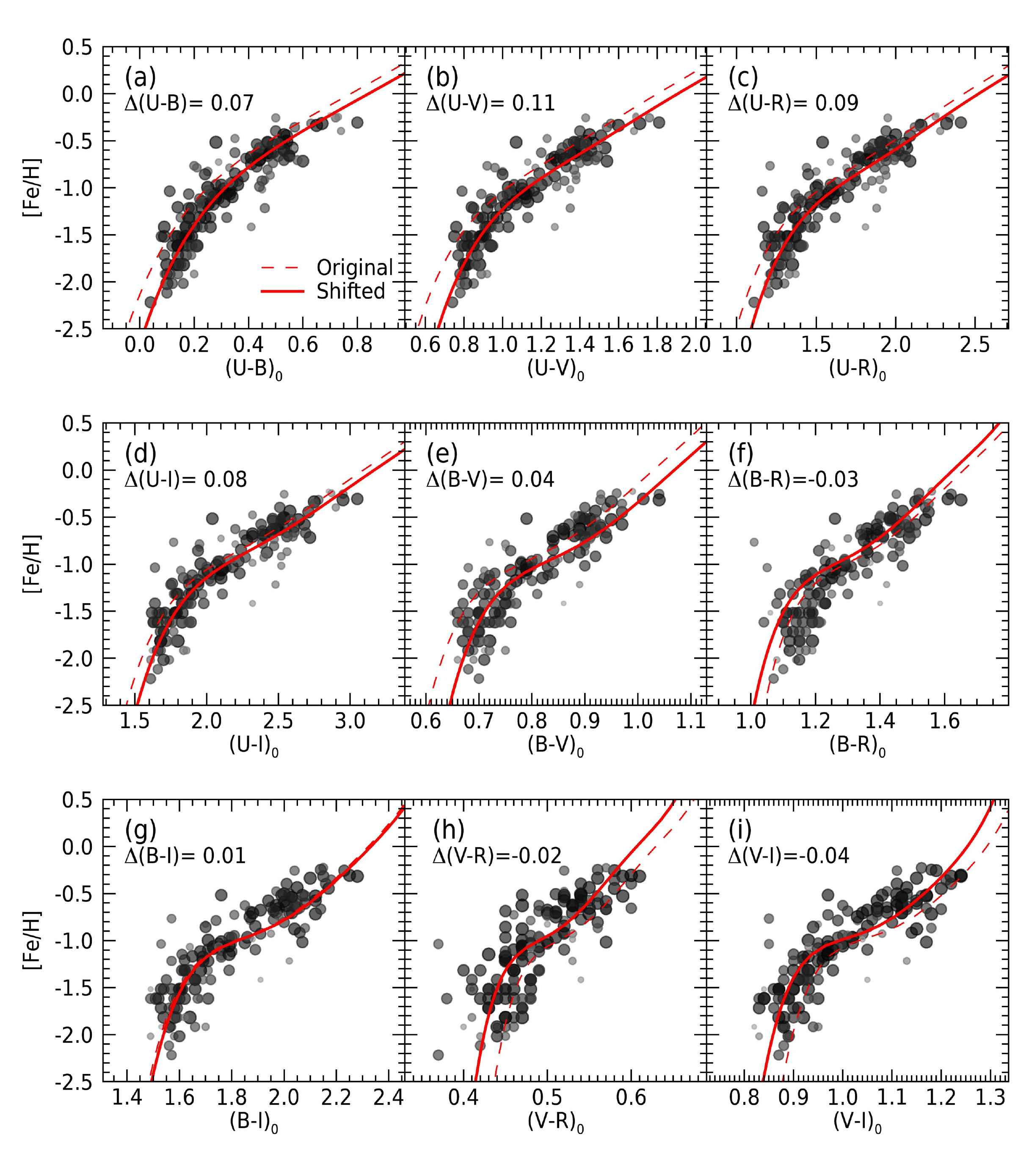}
\caption{
CMRs of NGC 5128 GCs. The size and opacity of each symbol are inversely proportional to the quadratic sum of observational uncertainties calculated after normalization. The red dashed and solid lines respectively represent the original 12.5 Gyr YEPS model predictions and the models shifted by the offset value denoted on the top-left corner of each panel. The overall shape of the CMRs is in good agreement with the observations and models for all colors, in that the slope of GC CMRs is steep in the blue region, decreases with redder, and has a quasi-inflection point along $BVRI$ CMRs [panels (d)--(i)].
\label{fig:cmr_5128}}
\end{figure}

Figure~\ref{fig:cmr_5128} shows the CMRs of NGC 5128 GCs in nine color combinations. 
The size and opacity of each symbol are set to be inversely proportional to the quadratic sum of errors calculated after normalization.
The red dashed and solid lines respectively represent the original 12.5 Gyr YEPS model predictions and the models shifted by the offset value denoted on the top-left corner of each panel.
We verify that the overall shape of the CMRs agrees well between observations and models for all the colors.
The scatters may result from both observational uncertainties and the spread of intrinsic properties of GCs such as age and $\alpha$-element abundance. 
It is a common feature that the slope of GC CMRs is steep in the blue region and decreases as the color becomes redder, which agrees with the features of the YEPS model prediction.
The characteristics are also consistent with previously suggested nonlinear empirical CMRs such as broken-line CMRs \citep[e.g.,][]{Peng06, Usher12} and high-order polynomial CMRs \citep[e.g.,][]{Blakes10, Vanderbeke14}.

The presence of a quasi-inflection point along $BVRI$ CMRs is an essential feature in view of the color bimodality origin debate.
The observations, although inconclusive, give hints of inflection along the $BVRI$ CMRs [panels (d)-(i)].
A paucity of GCs occurs at the vicinity of the quasi-inflection point where the color changes faster than the metallicity. 
The paucity near the inflection point is evident along the color axis, which makes CDs bimodal even from the unimodal MD (Paper I).
We will investigate this effect further in the next session.

\begin{figure}
\epsscale{1.2}
\plotone{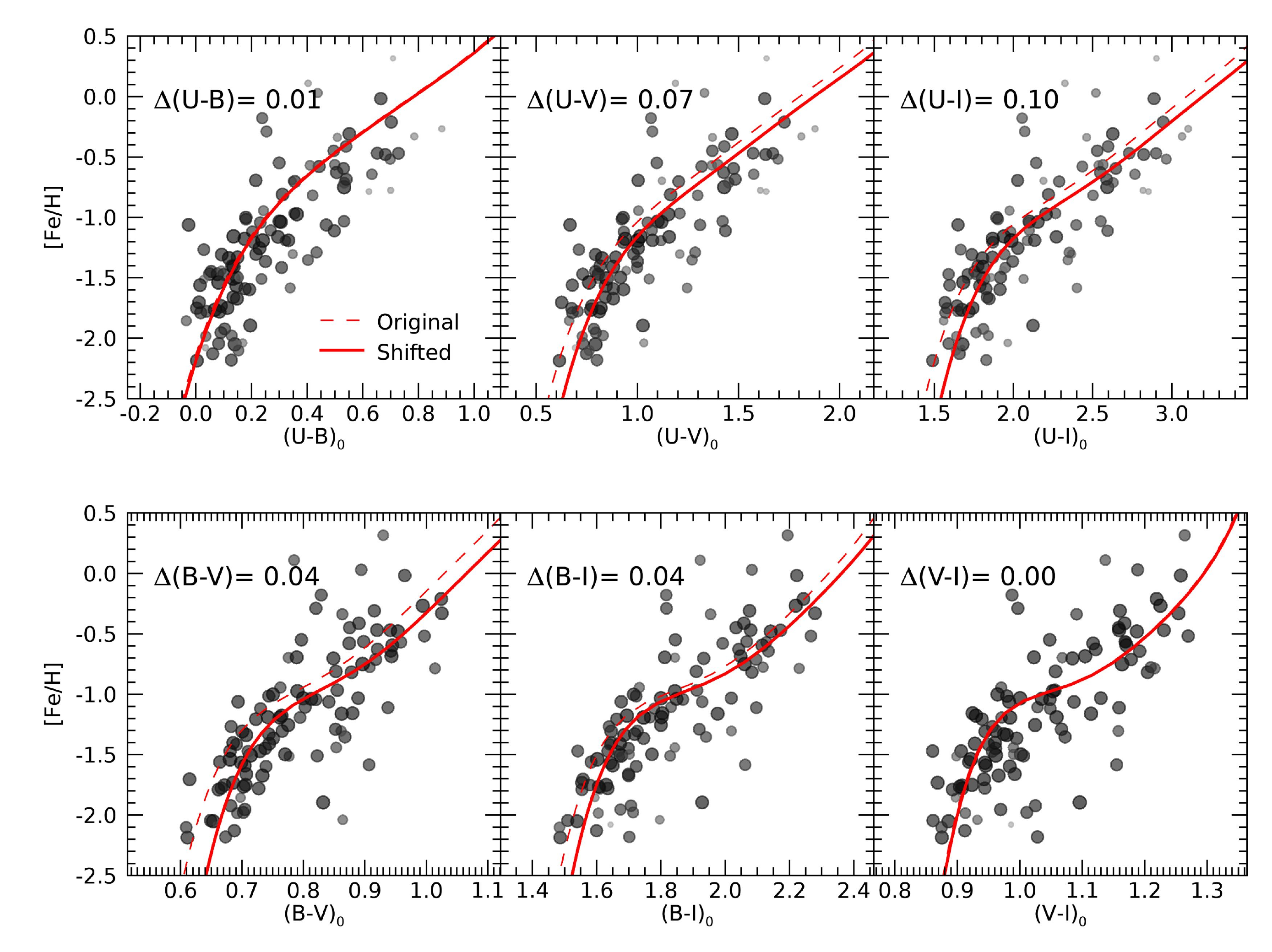}
\caption{
Same as Figure~\ref{fig:cmr_5128}, but for NGC 4594 GCs. The scatters are rather larger compared to the NGC 5128 GC CMRs, but similar features are shown: the CMR slope is steeper in the blue and the slope-changing point appears to be consistent with the model prediction. 
\label{fig:cmr_4594}}
\end{figure}

Figure~\ref{fig:cmr_4594} shows the CMRs of NGC 4594 GCs for six color combinations. 
The symbols and lines are the same as in Figure~\ref{fig:cmr_5128}.
Although the scatters are rather larger compared to the NGC 5128 GC CMRs, there are similar features in that the CMR slope is steeper in the blue and the slope-changing point appears to be consistent with the model prediction.

\begin{figure}
\epsscale{1.2}
\plotone{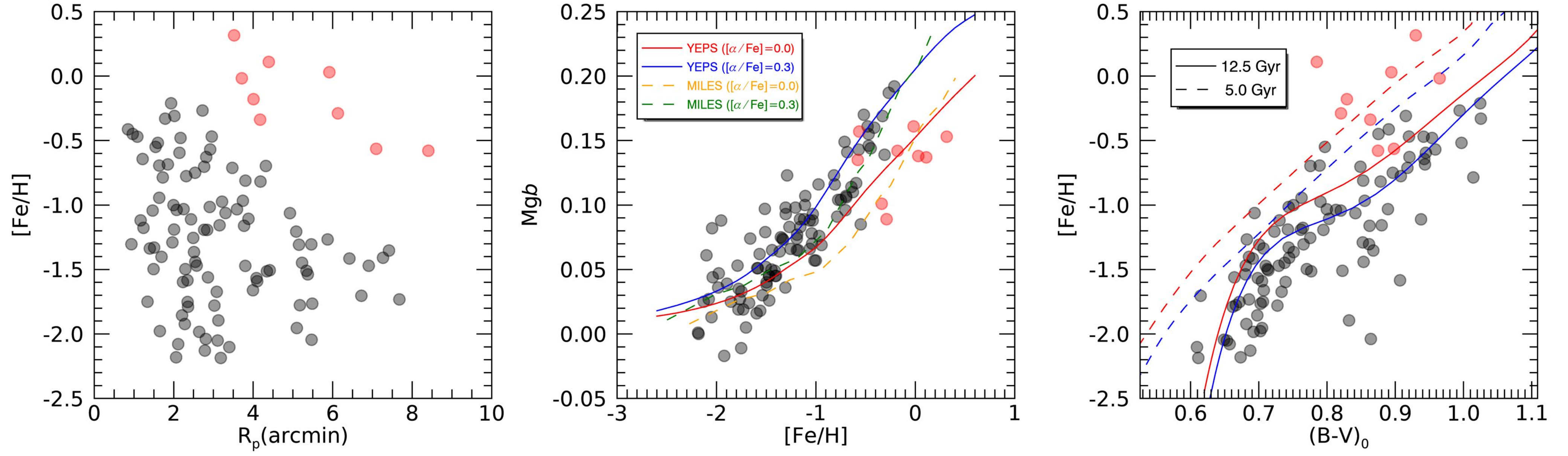}
\caption{
Elucidation of outliers in the CMR plane of NGC 4594 GCs.
To inspect the nature of the outliers, we present three plots 
showing GC metallicities as a function of galactocentric distance (left panel), 
the Mg$b$ versus [Fe/H] correlation along with 12.5 Gyr model predictions of the YEPS model (solid lines), 
and MILES model (dashed lines) for the different $\alpha$-element mixture (middle panel),
and the $B-V$ CMR along with the 5 Gyr and 12.5 Gyr YEPS model predictions 
for [$\alpha$/Fe] = 0.0 (blue lines) and 0.3 (red lines) (right panel).
In the left panel, the red circles indicate outer-halo, metal-rich GCs 
that do not follow the metallicity gradient.} 
We suspect that they are young GCs with a different formation and evolutionary history from most GCs in the galaxy, 
as they are less $\alpha$-element enhanced than other GCs with similar metallicities in the middle panel 
and are in the younger-age region of CMR in the right panel.
\label{fig:out_4594}
\end{figure}

Noticeably, there are outliers away from the model CMR predictions, particularly among red GCs. 
To scrutinize the nature of these GCs, in Figure~\ref{fig:out_4594}, we present their metallicities as a function of galactocentric distance (left), Mg$b$ vs. [Fe/H] along with model predictions for different $\alpha$-elements mixture (middle), and the $B-V$ CMR along with the 5 Gyr and 12.5 Gyr model predictions for [$\alpha$/Fe] = 0.0 and 0.3 (right).
For sanity check, the middle panel includes the MILES model predictions \citep{Vazdekis15} for [$\alpha$/Fe] = 0.0 and 0.3.
GC systems of massive galaxies usually exhibit a metallicity gradient \citep[e.g.,][]{Forbes18, Lee20}, in the sense that metal-rich GCs are more centrally concentrated than metal-poor GCs.  
In the left panel, the inner-halo GCs (R $<$ 3\arcmin)\footnote{The effective radius of NGC 4594 is 2\arcmin.6 \citep{Jardel11}.} show a large metallicity spread. 
The outer-halo GCs (R $>$ 3\arcmin) are divided into two groups, one of which appears to naturally follow the metallicity gradient while the other group has unexpectedly high metallicity given the radial distance. 
We mark the outer-halo, metal-rich GCs as red dots. 
The middle panel indicates that they are less enhanced in $\alpha$ elements than the other GCs with similar metallicities (at [Fe/H] = $-$0.5 $\sim$ 0.5).
In the right panel, most of the outer-halo, metal-rich GCs are located in the younger-age region of CMRs. 
As we mentioned in Section~\ref{sec:GCselect}, our sample could be contaminated by young GCs with intermediate/high metallicities.
We therefore suspect that these metal-rich outliers are young GCs with a history different from that of most GCs in the galaxy. For example, they might be created from metal-enriched gas in other galaxies through a relatively prolonged star formation and later accreted to the Sombrero.
For the metal-poor red outliers, on the other hand, no anomalous properties are found in terms of spatial distribution, radial velocities, metallicity gradient, and $\alpha$-enhancement.

\section{Nonlinear Metallicity-to-color Conversion and Color Bimodality} \label{sec:MDCD}

This section focuses on the effect of the nonlinear GC CMR on the metallicity-to-color transformation. 
For the GC samples of NGC 5128 and NGC 4594 that are crossmatched between photometry and spectroscopy, we present the MD and CDs for each galaxy and analyze the morphologies of the distributions in terms of the degree of bimodality.
We then demonstrate that the nonlinear metallicity-to-color conversion, combined with the observed GC MDs of NGC 5128 and NGC 4594, reproduces the observed bimodal CDs.

Figures~\ref{fig:md_cd_n5128} and~\ref{fig:md_cd_n4594} present the MD and CDs for NGC 5128 GCs and NGC 4594 GCs, respectively.
The MD and CDs in the lower panels of each figure are for the identical crossmatched GC samples to those used for the CMR nonlinearity test performed in Section~\ref{sec:CMRs}. 
The upper panels of each figure show, for reference, the CDs of photometrically selected old GC candidates, which have the same properties as those of the crossmatched GC samples in terms of the magnitude range and the radial coverage on the sky.
Each histogram is smoothed using a nonparametric kernel density estimate with the bandwidth determined by the mean observational error. To mitigate spiky features in the histogram that arises from the small number statistics, we use the same bandwidth for both photometrically selected GCs (upper panel) and crossmatched GCs (lower panel). 
We note that, for the MD of NGC 4594 GCs, \citet{Alv11} mentioned that their metallicity uncertainties might be underestimated, and the bandwidth is taken as twice of the mean observational error.
The smoothed histograms are denoted as red lines for MDs and green lines for CDs.
For NGC 5128 GCs, the MD can be described as a broad, unimodal distribution, but the CDs appear clearly bimodal, with the detailed shape being different depending on colors. 
The MD has a metal-rich peak but the CDs have prominent blue peaks. 
Given that metal-rich GCs are red, it is interesting that the shape of the MD is deformed dramatically after conversion into the CDs.
For NGC 4954 GCs, the MD is unimodal and slightly tilted toward the metal-poor side, and the CDs exhibit more discernible bimodality and more prominent blue peaks and dips compared to the MD.
In agreement with NGC 5128 GCs, the degree of CD bimodality of NGC 4594 GCs varies depending on color. 
This is consistent with the result for the GC systems of M84, M87, and NGC 1399 (Papers II and IV; \citealt{HSKim13}).

To evaluate the degree of bimodality of the MDs and CDs, we calculate the bimodality coefficient (BC; \citealt{SAS90}), defined as 
\begin{equation}
{BC = \frac{m_{3}^{2} + 1}{m_{4} + 3\cdot\frac{(n-1)^{2}}{(n-2)(n-3)}} \,\,,}
\end{equation}
where $n$ refers to the number of samples, and $m_{3}$ and $m_{4}$ are respectively the skewness and excess kurtosis of the distribution corrected for sample bias.
The BC exploits the skewness and the kurtosis of a distribution, which measure asymmetry and central clustering of the distribution, respectively.
The basic idea is that a bimodal distribution generally has higher skewness and lower kurtosis than a simple normal distribution. 
The higher BC indicates stronger bimodality, and the lower BC indicates weaker bimodality.
Readers are referred to \citet{Knapp07}, \citet{Pfister13}, and references therein for further details.
Table~\ref{tab:statis} presents the BCs for MD and CDs of GCs in NGC 5128 and NGC 4594.
The BCs alike indicate that the CDs are more bimodal than the MD for both galaxies.
This is consistent with the prediction of the nonlinear CMR theory, in which the inflection point along CMRs evokes or enhances bimodality by creating a dip in the CDs.
The BCs also suggest that the degree of CD bimodality varies with color.
Consistent with the visual impression above, both indices for $U-B$ CDs of the two galaxies designate weaker bimodality compared to the other CDs, except for the BC value for NGC 4594 GCs\footnote{\citet{Pfister13} show that the BC can give erroneous results in that highly skewed unimodal distributions have larger BC values (i.e., more bimodal) than skewed bimodal distributions \citep[see also][]{Kang19}.}.

We also carry out two statistical tests; Hartigans’ dip test \citep{Hartigan85} and the bimodality test using Gaussian Mixture Model (GMM) procedure. The former test is widely used for assessing the unimodality of a distribution by computing the maximum difference between the input distribution and the unimodal (uniform) distribution. The latter test assumes that the input distribution consists of multiple Gaussian components and constructs the GMMs that best describe the input distributions by maximum likelihood using the expectation--maximization algorithm. We use the $diptest$ \citep{Maechler21} and $mclust$ \citep{Scrucca16} packages implemented in R \citep{R22} to perform the dip test and the GMM test (we refer to it as GMM$_{R}$) respectively. In the GMM$_{R}$ test, we select the `V' model that assumes the Gaussian components can have different variances. We also conduct a revised GMM test provided by \citet[we refer to it as GMM$_{\rm MG}$]{Muratov10}, that employs three different approaches (the ratio of the likelihood, the peak separation, and the kurtosis) to get a more robust measure of bimodality.

All the tests provide the probability (i.e., $p$-value) for the null hypothesis that a distribution is unimodal. We note that the two tests occasionally give erroneous results. For example, the dip test hardly distinguishes between multimodal and unimodal distribution when the sample size is small and/or the significance level is set to be low \citep{Kang19}. The GMM$_{\rm MG}$ test often falsely indicates bimodal when the true distribution is unimodal but skewed because of its sensitivity to the assumption of Gaussian modes \citep{Muratov10}. As these issues are beyond the scope of this paper, we focus on the {\it relative} comparison of the $p$-values for the MDs and CDs for each test.

Table~\ref{tab:tests} provides the $p$-values from the dip test and the GMM tests for the MDs and CDs of NGC 5128 and NGC 4594 GCs. We remind that the smaller the $p$-values, the stronger the bimodality. The dip test clearly shows that the MDs are closest to a unimodal distribution, and the CDs are relatively close to bimodal distribution for both galaxies. The GMM$_{\rm MG}$ test gives three probabilities of a unimodal distribution based on maximum likelihood, peak separation, and kurtosis. They show that, with a few exceptions, the GC MDs for both galaxies are closer to unimodal distributions than their GC CDs. The GMM$_{R}$ test shows that, for NGC 5128, the $p$-value for MD is slightly larger, but a bimodal distribution is preferred over a unimodal distribution for all distributions. The $p$-values for NGC 4594 indicate that the MD is close to unimodal and the CDs are close to bimodal.

Figure~\ref{fig:gmm_r} presents Bayesian information criterion (BIC; \citet{Schwarz78}) and integrated complete data likelihood criterion (ICL; \citet{Biernacki00}) values in the GMM$_{R}$ test as a function of the number of Gaussian components $K$. 
BIC and ICL are usually used to select the optimal model in a way that $K$-components GMM with the highest BIC/ICL value is preferred.
BIC tends to select the number of mixture components, whereas ICL is more useful for selecting the number of discrete groups in the data \citep{Biernacki00}.  Given the properties of our data, it is more appropriate to focus on BIC rather than on ICL.
Consistent with the $p$-values, BIC values are highest at $K=2$ for all distributions except for the MD of NGC 4594 GCs. Note that, for the MD of NGC 5128, the difference in BIC values between $K=1$ and $K=2$ is $\sim$1.5. The difference between BIC values less than 2 is considered a “barely worth mentioning” difference \citep{Raftery95}. Thus, the GMM$_{R}$ test indicates that the MDs are relatively unimodal in both galaxies, as opposed to the CDs being bimodal.

The statistical tests corroborate that the distributions are significantly deformed when GC metallicities are converted to GC colors. The large variation in $p$-values among different colors (e.g., the $p$-value of the dip test for the NGC 4594 GC system) also indicates a significant change in their degree of bimodality due to the color-dependent shape of the CMR.

In Figures~\ref{fig:rep_5128} and~\ref{fig:rep_4594}, we perform Monte Carlo simulations aiming at the CDs of NGC 5128 and NGC 4594 GCs to validate the nonlinear CMR effect on the conversion of MDs into CDs. 
The top row of the figures shows the observed GC CMRs as shown in Figures~\ref{fig:cmr_5128} and~\ref{fig:cmr_4594}, overplotted with the YEPS model predictions. The observed MDs are rotated and given on the $y$-axis.
The second rows show the observed color--magnitude diagrams of the crossmatched GCs (black dots) and the photometrically selected GCs (gray dots). Observational uncertainties as a function of $M_{V}$ are denoted by error bars on the left of each panel.
The third and the fourth rows show the CDs of the crossmatched GCs and photometrically selected GCs, respectively (same as the Figures~\ref{fig:md_cd_n5128} and~\ref{fig:md_cd_n4594}). 
The bottom row of the figures shows the simulated GC CDs.
We generate 100,000 model GCs with metallicities derived from the Kernel density estimation of the observed MDs. 
We then transform the metallicity into colors using the CMRs predicted from the YEPS model, taking into account the observational uncertainties.
For each color, we randomly select the same number of model GCs with the observed GCs 10,000 times and calculate the 1$\sigma$ distribution ranges, which are shown by the shaded bands.

The most conspicuous feature of the observed CDs is the prominent blue peak. 
Our simulations with nonlinear CMRs can successfully reproduce the observed prominent blue peak that cannot be explained by the conventional linear CMR scheme given the unimodal MD with a metal-rich peak.
Metal-poor GCs are projected through the steeper part of the CMR into a narrow color region, producing the sharp blue peak in the CD. 
By contrast, metal-rich GCs are projected through the shallower part of the CMR into a wider color region, resulting in an attenuated red peak in the CD. 
This even evokes a reversal of peak intensity for the NGC 5128 GCs.

The simulations show that projections on nonlinear CMRs with inflection points cause bimodal CDs from the observed MDs close to unimodal distributions. 
For the $B-V$, $B-I$, and $V-I$ colors, the inflection point in the CMR causes a dip between blue and red peaks, which makes a discernible bimodal feature in the CDs. 
On the other hand, for the $U-B$ color, the shape of the CD shows a skewed Gaussian without bimodality since there is no obvious inflection point in the $U-B$ CMR. 
As a consequence, the simulation results for $U-B$, $B-V$, $B-I$, and $V-I$ are consistent with the corresponding observations.

\section{Summary and Discussion} \label{sec:discuss}

In this study, we have examined the nonlinearity of GC CMRs and its effect on the conversion of MD into CDs. 
We have used multiband photometric data combined with spectroscopic data for NGC 5128 and NGC 4594 GCs. 
We have demonstrated clear nonlinearity of the old GC CMRs for both galaxies, which is in good agreement with the model predictions. 

Our visual inspection and statistical tests confirm the significant changes in the morphology of the CDs compared to the MD and even between the CDs. 
The most remarkable findings for NGC 5128 are that the metal-rich peak of the GC MD is significantly reduced into a weak red bump in the CDs, and that strong blue peaks unexpectedly emerge in the ($U$-band-free) CDs. 
Also, a distinct dip that is not obvious in the MD appears in between the blue and red peaks in the CDs. 
Similarly, for NGC 4594 GCs, the blue peaks of the CDs are sharper compared to the corresponding metal-poor peak in the MD, and the dips are clearly visible in the CDs, making the CDs bimodal. 
We have demonstrated through Monte Carlo simulations that all of these features arise from the nonlinear conversion from the MDs to the CDs.

\begin{figure}
\epsscale{1.2}
\plotone{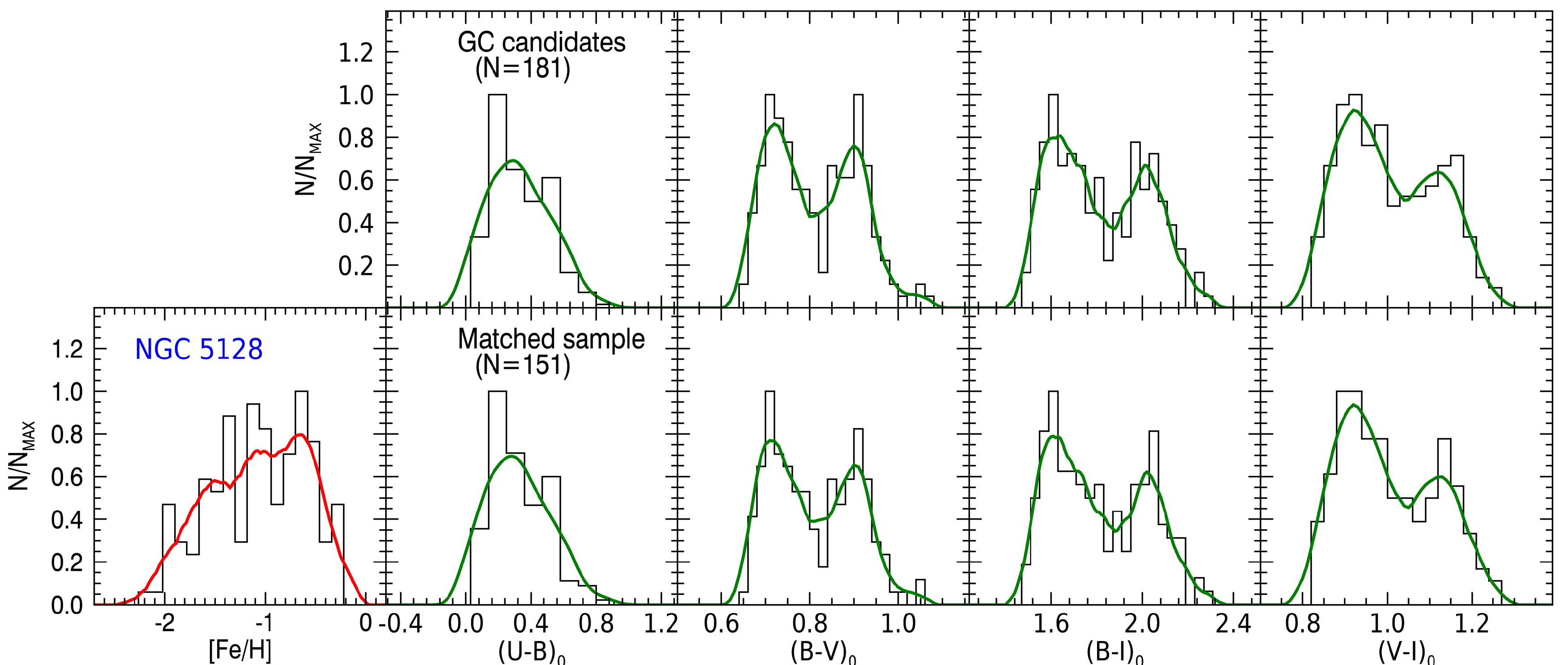}
\caption{
MD and CDs for GCs in NGC 5128. Each histogram is smoothed using a nonparametric kernel density estimate with the bandwidth corresponding to the mean observational error. The smoothed histograms are denoted as red lines for MDs and green lines for CDs. Upper panels: CDs for the GC candidates selected only from the photometric data set. Old GC candidates are selected from the photometric data of \citet{Peng06} and \citet{Wood07} using the same criteria as the two-color diagrams, magnitude range, and radial extent used for the crossmatched GC sample. Lower panels: MD and CDs for the GCs crossmatched between photometry and spectroscopy.
\label{fig:md_cd_n5128}}
\end{figure}

\begin{figure}
\epsscale{1.2}
\plotone{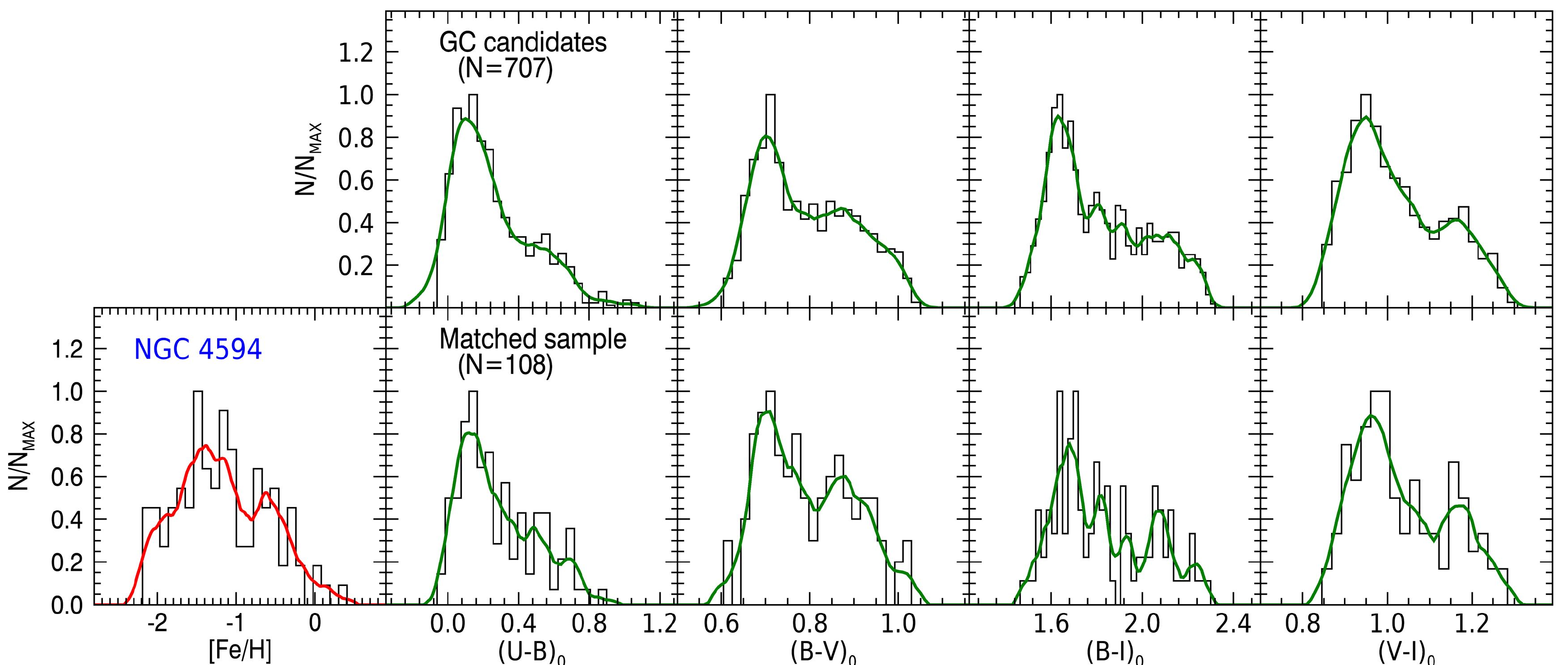}
\caption{
MD and CDs for GCs in NGC 4594. Each histogram is smoothed using a nonparametric kernel density estimate with the bandwidth corresponding to the mean observational error. The kernel bandwidth of MD is set to be twice the mean metallicity error since \citet{Alv11} mentioned that the metallicity error they provide may be underestimated. The smoothed histograms are denoted as red lines for MDs and green lines for CDs. Upper panels: CDs for the GC candidates selected only from the photometric data set. Old GC candidates are selected from our photometric data using the same criteria as the two-color diagrams, magnitude range, and radial extent used for the crossmatched GC sample. Lower panels: MD and CDs for the GCs crossmatched between photometry and spectroscopy.
\label{fig:md_cd_n4594}}
\end{figure}

\begin{figure}
\epsscale{1.0}
\plotone{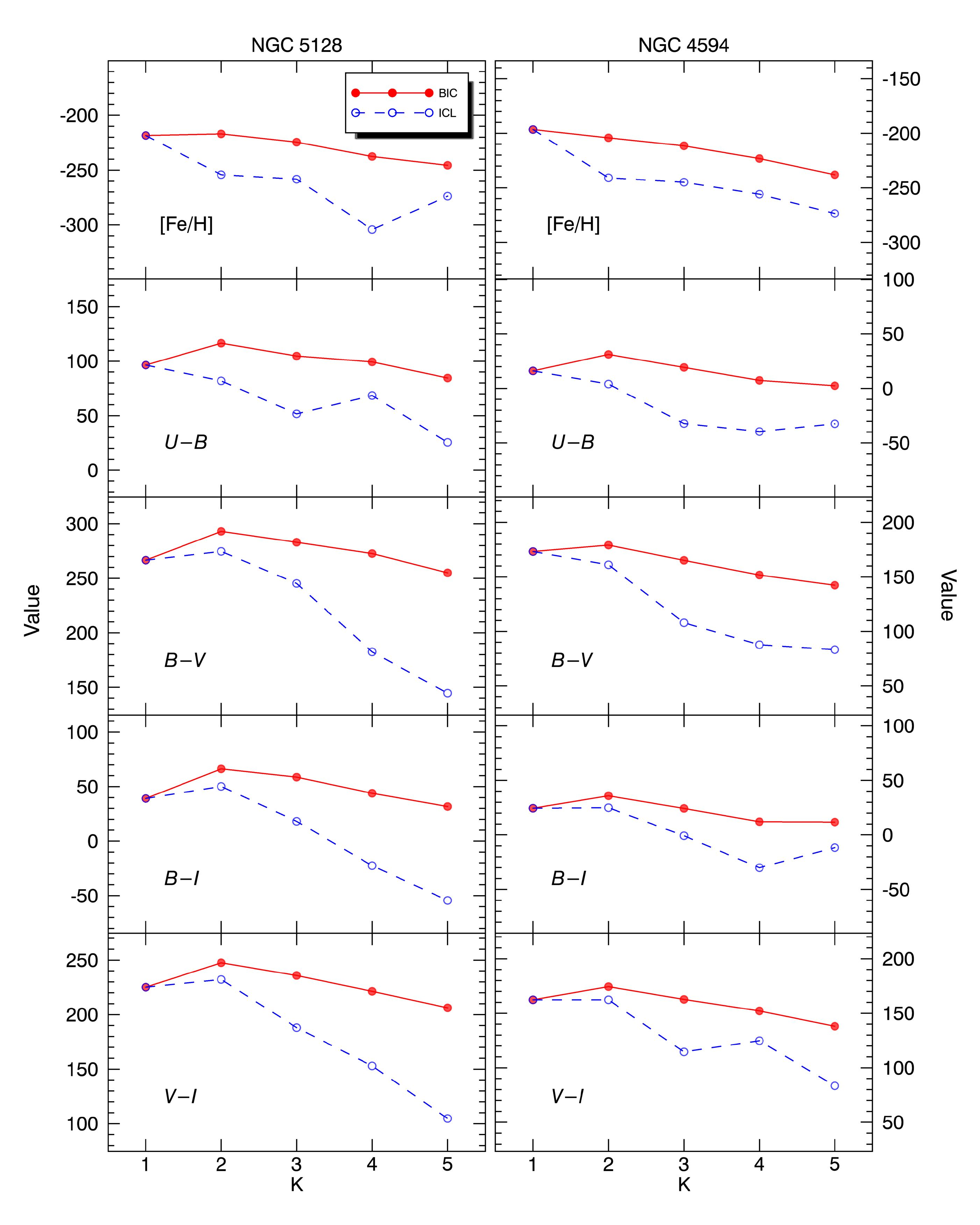}
\caption{The BIC (solid red circle) and ICL (open blue circle) values as functions of the number of Gaussian components $K$. The BIC and ICL values are used as criteria for the selection of the best mixture model to fit the data in the GMM$_{R}$ test. The models with the highest BIC/ICL values are preferred. Note that BIC tends to select the number of Gaussian components, whereas ICL is better suited for selecting discrete groups from data.
\label{fig:gmm_r}}
\end{figure}

\begin{figure}
\epsscale{1.2}
\plotone{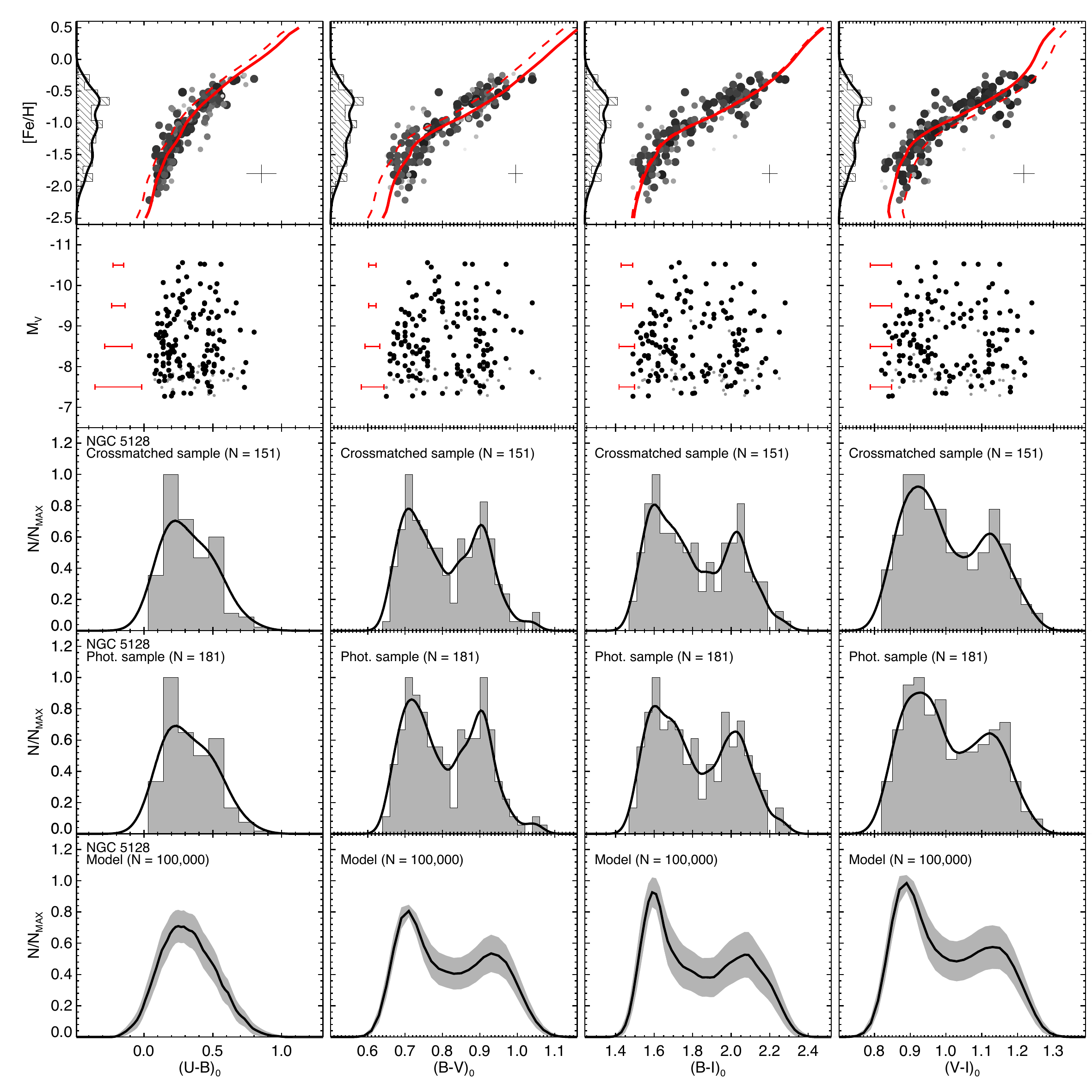}
\caption{Monte Carlo simulations for the CDs of NGC 5128 GCs. 
Top row: same as the CMRs in Figure~\ref{fig:cmr_5128}. The error bars indicate the typical errors in [Fe/H] and colors.
The observed MD is rotated and given on the $y$-axes, and the black solid line shows a kernel density estimate with a bandwidth set to the mean observational error.
Second row: the observed color-magnitude diagrams for the GCs. Black and gray dots represent the crossmatched and photometrically selected GCs, respectively. Observational uncertainties are shown by error bars as a function of $M_{V}$.
Third row: observed CDs of the crossmatched GCs (see Figure~\ref{fig:md_cd_n5128}).
Fourth row: observed CDs of the photometrically selected GCs. 
Bottom row: simulated CDs of model GCs. We generate $10^{5}$ model GCs with metallicities derived from the kernel density estimate of the observed MDs and convert the metallicities into colors via the model CMRs presented in the top panel taking into account the observational errors. Given that the number of observed GCs is much smaller, we estimate the CD of 151 model GCs (solid line) and its 1$\sigma$ distribution range (gray shaded bands) from the 10,000 bootstrap sampling. 
\label{fig:rep_5128}}
\end{figure}

\begin{figure}
\epsscale{1.2}
\plotone{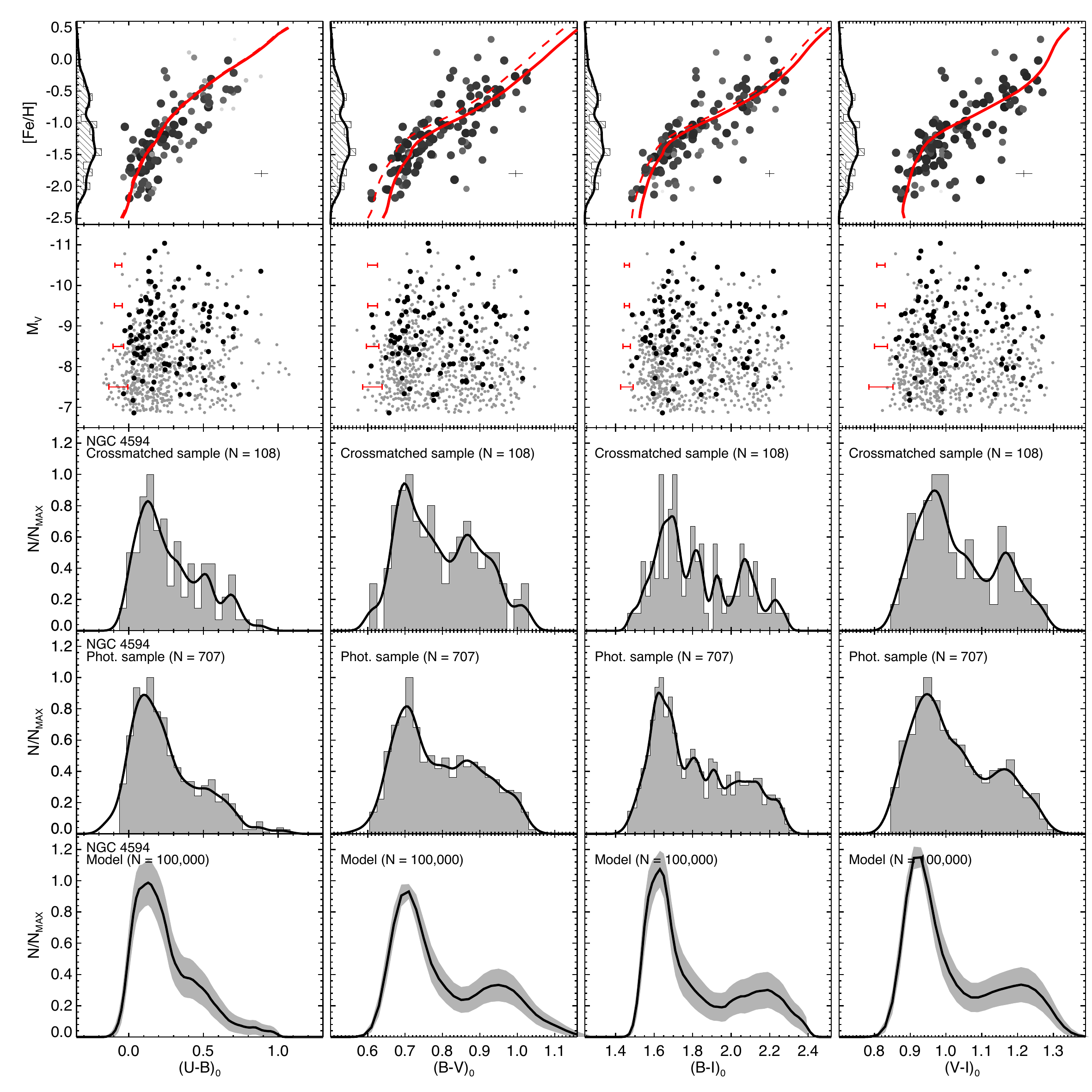}
\caption{
Monte Carlo simulations for the CDs of NGC 4594 GCs. 
Top row: same as the CMRs in Figure~\ref{fig:cmr_4594}. The error bars indicate the typical errors in [Fe/H] and colors. We note that \citet{Alv11} remark that their metallicity errors might be underestimated. The observed MD is rotated and given on the $y$-axes, and the black solid line shows a kernel density estimate with a bandwidth set to twice the mean observational error.
Second row: the observed color-magnitude diagrams for the GCs. Black and gray dots represent the crossmatched and photometrically selected GCs, respectively. Observational uncertainties are shown by error bars as a function of $M_{V}$.
Third row: observed CDs of the crossmatched GCs (see Figure~\ref{fig:md_cd_n4594}).
Fourth row: observed CDs of the photometrically selected GCs. 
Bottom row: simulated CDs of model GCs. We generate $10^{5}$ model GCs with metallicities derived from the kernel density estimate of the observed MDs and convert the metallicities into colors via the model CMRs presented in the top panel taking into account the observational errors. Given that the number of observed GCs is much smaller, we estimate the CD of 108 model GCs (solid line) and its 1-$\sigma$ distribution range (gray shaded bands) from the 10,000 bootstrap sampling. 
\label{fig:rep_4594}}
\end{figure}

The CDs of photometrically selected old GCs are very similar to those of the GC samples crossmatched between photometry and spectroscopy in both galaxies (Figures~\ref{fig:md_cd_n5128} and~\ref{fig:md_cd_n4594}). 
This implies that the MDs observed in this study are representative MDs of old GCs that are detectable at the corresponding galactocentric distance range. 
The shape of the GC MDs in both galaxies is broad and unimodal. 
This is a natural expectation in the $\Lambda$ cold dark matter paradigm, where large galaxies such as NGC 5128 and NGC 4954 form through the hierarchical merging of numerous protogalaxies. 

It is interesting to note that for NGC 5128, the shape of the GC MD is fairly consistent with that of the halo field stars' MD (Figure 1 of \citealt{Rejkuba11}) and closely matches that of the MD inferred from the inverse transformation of CDs using nonlinear CMRs (Figure 6 of \citealt{Yoon11a}).
In the nonlinear CMR scenario, there is no discrepancy in MDs between GCs and halo field stars, and no further explanation for the excess of blue GCs is required.
The readers are referred to \citet{Yoon11a} for an in-depth discussion about solving the apparent discrepancy between GC MDs and halo stellar MDs.

Many studies on GC bimodality begin by dividing GCs into two groups in CDs---blue and red GCs. This is because, in CDs, GCs are distinctively divided into two groups by the dip. 
However, in our theory, the clear division in CDs do not necessarily indicate the presence of the two groups in an MD in a single galaxy. 
The crossmatched GC samples for NGC 5128 and NGC 4594 show that the CDs are bimodal but the corresponding MDs are closer to unimodal than bimodal. 
A crucial point of GC color bimodality is that it is a universal phenomenon commonly observed in massive early-type galaxies \citep{Brodie06}. 
The detailed formation process of these galaxies varies from case to case, but there should be an essential common process to give rise to the phenomenon. 
In this respect, the nonlinear CMR scenario has its own strength in that the projection effect can simply solve the ubiquity of GC color bimodality. 
This study provides further direct observational evidence (see also Papers I--X) favoring the projection effect that significantly deforms MDs into various CD morphologies. 
Therefore, we conclude that the nonlinear CMR effect is the basic cause of the color bimodality phenomenon, and the detailed characteristics of CDs vary depending on the formation history of each galaxy.

\begin{acknowledgments}
H.-S.K. acknowledges financial support from the Basic Science Research Program (No. 2019R1A2C2086290) through the National Research Foundation of Korea (NRF). 
S.-J.Y. acknowledges support from the Mid-career Researcher Program (No. 2019R1A2C3006242) through the NRF.
S.I.H. acknowledges the support provided by the Basic Science Research Program through the NRF funded by the Ministry of Education (No. 2020R1I1A1A01052358) and by the NRF grant funded by the Korean government (MSIT) (No. 2021R1A2C1004117).

\end{acknowledgments}

\clearpage
\begin{deluxetable}{lccrccccc}
\tablecolumns{9}
\tabletypesize{\small}
\tablecaption{$UBVRI$ Photometry and Spectroscopic Metallicity of Crossmatched GC Sample in NGC 5128}
\tablewidth{0pt}
\tablehead{\colhead{ID} & 
 \colhead{R.A. (J2000)} & 
 \colhead{Decl. (J2000)} & 
 \colhead{[Fe/H]\tablenotemark{a}} & 
 \colhead{$U_{0}$\tablenotemark{b}} & 
 \colhead{$B_{0}$\tablenotemark{b}} & 
 \colhead{$V_{0}$\tablenotemark{b}} & 
 \colhead{$R_{0}$\tablenotemark{b}} & 
 \colhead{$I_{0}$\tablenotemark{b}}
}
\startdata
   1 & 13:23:58.76 & --43:01:35.11 & --0.52 $\pm$ 0.08 & 19.91 $\pm$ 0.06 & 19.44 $\pm$ 0.02 & 18.54 $\pm$ 0.01 & 18.00 $\pm$ 0.01 & 17.37 $\pm$ 0.03 \\
   2 & 13:23:58.58 & --42:57:17.00 & --2.12 $\pm$ 0.17 & 20.03 $\pm$ 0.06 & 20.09 $\pm$ 0.02 & 19.52 $\pm$ 0.01 & 19.15 $\pm$ 0.01 & 18.75 $\pm$ 0.03 \\
   3 & 13:23:59.61 & --42:55:19.40 & --2.42 $\pm$ 0.48 & 20.01 $\pm$ 0.06 & 20.08 $\pm$ 0.02 & 19.53 $\pm$ 0.01 & 19.17 $\pm$ 0.01 & 18.80 $\pm$ 0.03 \\
   4 & 13:24:05.98 & --43:03:54.70 & --0.69 $\pm$ 0.18 & 21.04 $\pm$ 0.14 & 20.65 $\pm$ 0.03 & 19.79 $\pm$ 0.01 & 19.30 $\pm$ 0.01 & 18.78 $\pm$ 0.03 \\
   5 & 13:24:28.44 & --42:57:52.90 & --1.92 $\pm$ 0.24 & 20.59 $\pm$ 0.10 & 20.48 $\pm$ 0.03 & 19.79 $\pm$ 0.01 & 19.33 $\pm$ 0.01 & 18.91 $\pm$ 0.03 \\
   6 & 13:24:29.23 & --43:08:36.59 & --0.81 $\pm$ 0.13 & 21.27 $\pm$ 0.19 & 20.80 $\pm$ 0.04 & 19.90 $\pm$ 0.01 & 19.34 $\pm$ 0.01 & 18.80 $\pm$ 0.03 \\
   7 & 13:24:29.72 & --43:02:06.48 & --0.73 $\pm$ 0.27 & 20.68 $\pm$ 0.11 & 20.49 $\pm$ 0.03 & 19.74 $\pm$ 0.01 & 19.23 $\pm$ 0.01 & 18.77 $\pm$ 0.03 \\
   8 & 13:24:34.62 & --43:12:50.51 & --1.22 $\pm$ 0.34 & 21.17 $\pm$ 0.17 & 20.71 $\pm$ 0.04 & 19.82 $\pm$ 0.01 & 19.29 $\pm$ 0.01 & 18.69 $\pm$ 0.03 \\
   9 & 13:24:43.57 & --43:08:43.20 & --1.72 $\pm$ 0.11 & 19.11 $\pm$ 0.04 & 19.07 $\pm$ 0.01 & 18.45 $\pm$ 0.01 & 18.04 $\pm$ 0.01 & 17.66 $\pm$ 0.03 \\
  10 & 13:24:45.35 & --42:59:33.50 & --1.92 $\pm$ 0.15 & 19.38 $\pm$ 0.06 & 19.43 $\pm$ 0.02 & 18.98 $\pm$ 0.01 & 18.64 $\pm$ 0.01 & 18.13 $\pm$ 0.03 \\
  11 & 13:24:45.76 & --43:02:24.48 & --0.26 $\pm$ 0.09 & 21.32 $\pm$ 0.25 & 20.60 $\pm$ 0.04 & 19.56 $\pm$ 0.01 & 18.98 $\pm$ 0.01 & 18.37 $\pm$ 0.03 \\
  12 & 13:24:47.61 & --43:10:48.50 & --1.32 $\pm$ 0.25 & 20.34 $\pm$ 0.10 & 20.08 $\pm$ 0.03 & 19.32 $\pm$ 0.01 & 18.83 $\pm$ 0.01 & 18.37 $\pm$ 0.03 \\
  13 & 13:24:49.37 & --43:08:17.69 & --0.92 $\pm$ 0.31 & 20.93 $\pm$ 0.16 & 20.72 $\pm$ 0.04 & 19.99 $\pm$ 0.02 & 19.51 $\pm$ 0.01 & 19.02 $\pm$ 0.03 \\
  14 & 13:24:50.08 & --43:07:36.20 & --1.09 $\pm$ 0.42 & 20.92 $\pm$ 0.16 & 20.64 $\pm$ 0.04 & 20.11 $\pm$ 0.02 & 19.76 $\pm$ 0.02 & 19.28 $\pm$ 0.03 \\
  15 & 13:24:52.98 & --43:11:55.80 & --0.91 $\pm$ 0.31 & 20.63 $\pm$ 0.11 & 20.46 $\pm$ 0.03 & 19.77 $\pm$ 0.01 & 19.38 $\pm$ 0.01 & 18.92 $\pm$ 0.03 \\
... & ... & ... & ... & ... & ... & ... & ... & ... \\
... & ... & ... & ... & ... & ... & ... & ... & ... \\
... & ... & ... & ... & ... & ... & ... & ... & ... \\
\enddata
\label{tab:n5128gc}
\tablecomments{This table is available in its entirety in a machine-readable form in the online journal. A portion is shown here for guidance regarding its form and content.
\tablenotetext{a}{The metallicities are transformed from the empirical metallicity ([M/H]) provided by \citet{Beas08}.}
\tablenotetext{b}{The photometric data are from \citet{Wood07} and corrected for Galactic extinction (see text).}
}
\end{deluxetable}

\clearpage
\begin{deluxetable}{lccrcccc}
\tablecolumns{8}
\tabletypesize{\small}
\tablecaption{$UBVI$ Photometry and Spectroscopic Metallicity of Crossmatched GC Sample in NGC 4594}
\tablewidth{0pt}
\tablehead{\colhead{ID} &
 \colhead{R.A. (J2000)} &
 \colhead{Decl. (J2000)} & 
 \colhead{[Fe/H]\tablenotemark{a}} & 
 \colhead{$U_{0}$} & 
 \colhead{$B_{0}$} & 
 \colhead{$V_{0}$} & 
 \colhead{$I_{0}$}
}
\startdata
   1 & 12:40:10.48 & --11:45:20.56 & --0.58 $\pm$ 0.07 & 21.71 $\pm$ 0.02 & 21.27 $\pm$ 0.01 & 20.40 $\pm$ 0.01 & 19.28 $\pm$ 0.01 \\
   2 & 12:40:10.02 & --11:44:19.94 & --1.35 $\pm$ 0.08 & 22.64 $\pm$ 0.04 & 22.24 $\pm$ 0.01 & 21.37 $\pm$ 0.01 & 20.30 $\pm$ 0.01 \\
   3 & 12:40:08.50 & --11:43:24.49 & --1.41 $\pm$ 0.07 & 21.63 $\pm$ 0.02 & 21.32 $\pm$ 0.01 & 20.63 $\pm$ 0.01 & 19.68 $\pm$ 0.01 \\
   4 & 12:39:55.86 & --11:42:28.97 & --1.78 $\pm$ 0.06 & 22.28 $\pm$ 0.03 & 22.24 $\pm$ 0.01 & 21.58 $\pm$ 0.01 & 20.63 $\pm$ 0.01 \\
   5 & 12:40:05.67 & --11:42:18.26 & --1.31 $\pm$ 0.05 & 21.82 $\pm$ 0.02 & 21.72 $\pm$ 0.01 & 21.02 $\pm$ 0.01 & 20.08 $\pm$ 0.01 \\
   6 & 12:40:10.04 & --11:42:11.78 & --2.05 $\pm$ 0.05 & 22.41 $\pm$ 0.03 & 22.33 $\pm$ 0.01 & 21.68 $\pm$ 0.01 & 20.82 $\pm$ 0.01 \\
   7 & 12:40:11.71 & --11:41:56.65 & --1.30 $\pm$ 0.07 & 21.61 $\pm$ 0.02 & 21.39 $\pm$ 0.01 & 20.63 $\pm$ 0.01 & 19.66 $\pm$ 0.01 \\
   8 & 12:40:14.63 & --11:41:55.60 & --1.27 $\pm$ 0.06 & 22.52 $\pm$ 0.03 & 22.49 $\pm$ 0.01 & 21.80 $\pm$ 0.01 & 20.85 $\pm$ 0.02 \\
   9 & 12:40:12.63 & --11:41:29.59 & --1.45 $\pm$ 0.05 & 22.40 $\pm$ 0.03 & 22.34 $\pm$ 0.01 & 21.60 $\pm$ 0.01 & 20.64 $\pm$ 0.01 \\
  10 & 12:40:04.65 & --11:40:53.91 & --0.97 $\pm$ 0.05 & 22.91 $\pm$ 0.04 & 22.56 $\pm$ 0.01 & 21.71 $\pm$ 0.01 & 20.65 $\pm$ 0.02 \\
  11 & 12:40:01.36 & --11:40:52.32 &  0.32 $\pm$ 0.06 & 24.03 $\pm$ 0.10 & 23.32 $\pm$ 0.02 & 22.39 $\pm$ 0.02 & 21.12 $\pm$ 0.02 \\
  12 & 12:40:15.01 & --11:40:44.44 & --1.21 $\pm$ 0.05 & 22.40 $\pm$ 0.03 & 22.20 $\pm$ 0.01 & 21.47 $\pm$ 0.01 & 20.53 $\pm$ 0.01 \\
  13 & 12:40:07.23 & --11:40:25.56 & --1.03 $\pm$ 0.04 & 21.50 $\pm$ 0.02 & 21.20 $\pm$ 0.01 & 20.40 $\pm$ 0.01 & 19.40 $\pm$ 0.01 \\
  14 & 12:40:25.81 & --11:40:19.07 & --0.56 $\pm$ 0.05 & 23.32 $\pm$ 0.05 & 22.82 $\pm$ 0.01 & 21.92 $\pm$ 0.01 & 20.75 $\pm$ 0.02 \\
  15 & 12:40:01.23 & --11:40:07.90 & --2.13 $\pm$ 0.05 & 22.19 $\pm$ 0.03 & 22.13 $\pm$ 0.01 & 21.44 $\pm$ 0.01 & 20.53 $\pm$ 0.01 \\
... & ... & ... & ... & ... & ... & ... & ... \\
... & ... & ... & ... & ... & ... & ... & ... \\
... & ... & ... & ... & ... & ... & ... & ... \\
\enddata
\label{tab:n4594gc}
\tablecomments{This table is available in its entirety in a machine-readable form in the online journal. A portion is shown here for guidance regarding its form and content.
\tablenotetext{a}{The metallicities are from \citet{Alv11}.}}
\end{deluxetable}

\clearpage

\clearpage
\begin{deluxetable}{rrrrrrrrrrr}
\tablecolumns{11}
\tabletypesize{\small}
\tablecaption{Theoretical $UBVRI$ Colors and Metallicity for a 12.5 Gyr GC}
\label{tab:yeps_12.5}
\tablewidth{0pt}
\tablehead{\colhead{[Fe/H]} & \colhead{$U-B$} & \colhead{$U-V$} & \colhead{$U-R$} & \colhead{$U-I$} & \colhead{$B-V$} & \colhead{$B-R$} & \colhead{$B-I$} & \colhead{$V-R$} & \colhead{$V-I$} & \colhead{$R-I$}}
\startdata
 --2.5 & --0.052 &  0.552 &  0.991 &  1.438 &  0.603 &  1.043 &  1.489 &  0.440 &  0.886 &  0.446 \\
 --2.4 & --0.030 &  0.582 &  1.023 &  1.464 &  0.612 &  1.053 &  1.494 &  0.440 &  0.881 &  0.441 \\
 --2.3 & --0.015 &  0.601 &  1.042 &  1.484 &  0.617 &  1.058 &  1.500 &  0.441 &  0.883 &  0.442 \\
 --2.2 & --0.004 &  0.617 &  1.059 &  1.504 &  0.621 &  1.062 &  1.508 &  0.442 &  0.887 &  0.445 \\
 --2.1 &  0.006 &  0.630 &  1.073 &  1.523 &  0.624 &  1.067 &  1.516 &  0.443 &  0.892 &  0.449 \\
 --2.0 &  0.017 &  0.646 &  1.091 &  1.543 &  0.629 &  1.074 &  1.526 &  0.445 &  0.897 &  0.452 \\
 --1.9 &  0.029 &  0.665 &  1.112 &  1.566 &  0.636 &  1.083 &  1.537 &  0.448 &  0.901 &  0.454 \\
 --1.8 &  0.048 &  0.690 &  1.141 &  1.596 &  0.643 &  1.093 &  1.548 &  0.450 &  0.905 &  0.455 \\
 --1.7 &  0.067 &  0.719 &  1.173 &  1.630 &  0.652 &  1.107 &  1.563 &  0.454 &  0.911 &  0.457 \\
 --1.6 &  0.086 &  0.750 &  1.209 &  1.669 &  0.664 &  1.124 &  1.584 &  0.460 &  0.920 &  0.460 \\
 --1.5 &  0.105 &  0.782 &  1.247 &  1.713 &  0.677 &  1.142 &  1.608 &  0.466 &  0.931 &  0.465 \\
 --1.4 &  0.138 &  0.821 &  1.290 &  1.760 &  0.683 &  1.152 &  1.622 &  0.469 &  0.939 &  0.470 \\
 --1.3 &  0.171 &  0.867 &  1.342 &  1.821 &  0.696 &  1.171 &  1.650 &  0.475 &  0.954 &  0.478 \\
 --1.2 &  0.198 &  0.915 &  1.400 &  1.890 &  0.717 &  1.201 &  1.692 &  0.484 &  0.975 &  0.491 \\
 --1.1 &  0.220 &  0.966 &  1.464 &  1.971 &  0.747 &  1.244 &  1.751 &  0.497 &  1.004 &  0.507 \\
 --1.0 &  0.244 &  1.025 &  1.537 &  2.064 &  0.780 &  1.293 &  1.820 &  0.513 &  1.040 &  0.527 \\
 --0.9 &  0.282 &  1.099 &  1.629 &  2.178 &  0.817 &  1.347 &  1.896 &  0.529 &  1.079 &  0.549 \\
 --0.8 &  0.321 &  1.169 &  1.712 &  2.282 &  0.848 &  1.391 &  1.960 &  0.543 &  1.113 &  0.570 \\
 --0.7 &  0.366 &  1.241 &  1.797 &  2.387 &  0.875 &  1.431 &  2.021 &  0.556 &  1.146 &  0.590 \\
 --0.6 &  0.420 &  1.323 &  1.893 &  2.502 &  0.903 &  1.472 &  2.082 &  0.570 &  1.179 &  0.610 \\
 --0.5 &  0.475 &  1.402 &  1.985 &  2.612 &  0.928 &  1.510 &  2.137 &  0.582 &  1.210 &  0.627 \\
 --0.4 &  0.529 &  1.477 &  2.070 &  2.711 &  0.949 &  1.541 &  2.182 &  0.593 &  1.233 &  0.641 \\
 --0.3 &  0.586 &  1.556 &  2.158 &  2.809 &  0.969 &  1.571 &  2.223 &  0.602 &  1.253 &  0.651 \\
 --0.2 &  0.649 &  1.639 &  2.251 &  2.909 &  0.990 &  1.601 &  2.259 &  0.611 &  1.270 &  0.658 \\
 --0.1 &  0.717 &  1.727 &  2.347 &  3.009 &  1.010 &  1.630 &  2.292 &  0.620 &  1.282 &  0.662 \\
  0.0 &  0.784 &  1.814 &  2.442 &  3.106 &  1.030 &  1.658 &  2.322 &  0.628 &  1.292 &  0.663 \\
  0.1 &  0.847 &  1.897 &  2.534 &  3.198 &  1.049 &  1.687 &  2.351 &  0.638 &  1.302 &  0.664 \\
  0.2 &  0.903 &  1.971 &  2.619 &  3.282 &  1.068 &  1.716 &  2.379 &  0.648 &  1.311 &  0.663 \\
  0.3 &  0.952 &  2.040 &  2.699 &  3.362 &  1.088 &  1.747 &  2.410 &  0.658 &  1.322 &  0.663 \\
  0.4 &  1.000 &  2.108 &  2.777 &  3.444 &  1.108 &  1.776 &  2.444 &  0.669 &  1.336 &  0.668 \\
  0.5 &  1.072 &  2.201 &  2.877 &  3.553 &  1.128 &  1.804 &  2.481 &  0.676 &  1.352 &  0.676 \\
\enddata
\tablecomments{The spectrophotometric model data of the entire parameter space are available at \url{http://cosmic.yonsei.ac.kr/YEPS.htm}.}
\end{deluxetable}

\clearpage

\clearpage

\clearpage

\clearpage

\clearpage

\clearpage
\begin{deluxetable}{cccccc}
\tablecolumns{6}
\tablecaption{Bimodality Coefficient for Metallicity and Color Distributions of GCs in NGC 5128 and NGC 4594
\label{tab:statis}
}
\tablewidth{0pt}
\tablehead{\colhead{GC system} & \colhead{[Fe/H]} & \colhead{$U-B$} & \colhead{$B-V$} & \colhead{$B-I$} & \colhead{$V-I$}}
\startdata
  NGC~5128 & 0.487 & 0.513 & 0.536 & 0.568 & 0.559 \\
  NGC~4594 & 0.431 & 0.565 & 0.497 & 0.543 & 0.552 \\
\enddata
\end{deluxetable}

\begin{deluxetable}{ccccccc}
\tablecolumns{7}
\tabletypesize{\small}
\tablecaption{Results from Dip Test and GMM Test (in the Heteroscedastic Case) for Metallicity and Color Distributions of GCs in NGC 5128 and NGC 4594
\label{tab:tests}}
\tablewidth{0pt}
\tablehead{\colhead{GC system} & \colhead{Statistic} & \colhead{[Fe/H]} & \colhead{$U-B$} & \colhead{$B-V$} & \colhead{$B-I$} & \colhead{$V-I$}}
\startdata
\multirow{5}{4em}{NGC~5128} & $p(diptest)$ & 0.215 & 0.144 & 0.006 & 0.124 & 0.129 \\
                         {} & $p(GMM_{\rm MG}:\chi^{2})$ & 0.001 & 0.001 & 0.001 & 0.001 & 0.001 \\
                         {} & $p(GMM_{\rm MG}:DD)$       & 0.389 & 0.352 & 0.128 & 0.104 & 0.153 \\
                         {} & $p(GMM_{\rm MG}:kurt)$     & 0.001 & 0.034 & 0.001 & 0.001 & 0.001 \\
                         {} & $p(GMM_{R})$ & 0.004 & 0.001 & 0.001 & 0.001 & 0.001 \\
\hline
\multirow{5}{4em}{NGC~4594} & $p(diptest)$ & 0.912 & 0.850 & 0.792 & 0.492 & 0.364 \\
                         {} & $p(GMM_{\rm MG}:\chi^{2})$ & 0.157 & 0.001 & 0.001 & 0.001 & 0.001 \\
                         {} & $p(GMM_{\rm MG}:DD)$       & 0.421 & 0.476 & 0.249 & 0.147 & 0.136 \\
                         {} & $p(GMM_{\rm MG}:kurt)$     & 0.050 & 0.200 & 0.001 & 0.001 & 0.001 \\
                         {} & $p(GMM_{R})$ & 0.132 & 0.001 & 0.002 & 0.001 & 0.001 \\
\enddata
\end{deluxetable}

\clearpage

\clearpage

\clearpage


\begin{thebibliography}{}
\bibitem[Alves-Brito et al.(2011)]{Alv11} Alves-Brito, A., Hau, G.~K.~T., Forbes, D.~A., et al.\ 2011, \mnras, 417, 1823. doi:10.1111/j.1365-2966.2011.19368.x

\bibitem[Angelou et al.(2015)]{Ang15} Angelou, G.~C., D'Orazi, V., Constantino, T.~N., et al.\ 2015, \mnras, 450, 2423. doi:10.1093/mnras/stv770

\bibitem[Barmby \& Huchra(2000)]{Barmby00} Barmby, P. \& Huchra, J.~P.\ 2000, \apjl, 531, L29. doi:10.1086/312511

\bibitem[Bastian \& Lardo(2018)]{Bastian18} Bastian, N. \& Lardo, C.\ 2018, \araa, 56, 83. doi:10.1146/annurev-astro-081817-051839

\bibitem[Beasley et al.(2008)]{Beas08} Beasley, M.~A., Bridges, T., Peng, E., et al.\ 2008, \mnras, 386, 1443. doi:10.1111/j.1365-2966.2008.13123.x


\bibitem[Biernacki, Celeux, \& Govaert(2000)]{Biernacki00} Biernacki, C., Celeux, G. and Govaert, G.\ 2000, ITPAM, 22, 719. doi:10.1109/34.865189

\bibitem[Blakeslee et al.(2010)]{Blakes10} Blakeslee, J.~P., Cantiello, M., \& Peng, E.~W.\ 2010, \apj, 710, 51. doi:10.1088/0004-637X/710/1/51

\bibitem[Blakeslee et al.(2012)]{Blakes12} Blakeslee, J.~P., Cho, H., Peng, E.~W., et al.\ 2012, \apj, 746, 88. doi:10.1088/0004-637X/746/1/88
 
\bibitem[Brodie \& Huchra(1990)]{Brodie90} Brodie, J.~P. \& Huchra, J.~P.\ 1990, \apj, 362, 503. doi:10.1086/169288

\bibitem[Brodie \& Strader(2006)]{Brodie06} Brodie, J.~P. \& Strader, J.\ 2006, \araa, 44, 193. doi:10.1146/annurev.astro.44.051905.092441

\bibitem[Brodie et al.(2012)]{Brodie12} Brodie, J.~P., Usher, C., Conroy, C., et al.\ 2012, \apjl, 759, L33. doi:10.1088/2041-8205/759/2/L33

\bibitem[Cantiello \& Blakeslee(2007)]{Cantiello07} Cantiello, M. \& Blakeslee, J.~P.\ 2007, \apj, 669, 982. doi:10.1086/522110

\bibitem[Cantiello et al.(2014)]{Cantiello14} Cantiello, M., Blakeslee, J.~P., Raimondo, G., et al.\ 2014, \aap, 564, L3. doi:10.1051/0004-6361/201323272

\bibitem[Cardelli et al.(1989)]{Cardelli89} Cardelli, J.~A., Clayton, G.~C., \& Mathis, J.~S.\ 1989, \apj, 345, 245. doi:10.1086/167900

\bibitem[Chies-Santos et al.(2012)]{Chies12} Chies-Santos, A.~L., Larsen, S.~S., Cantiello, M., et al.\ 2012, \aap, 539, A54. doi:10.1051/0004-6361/201117169

\bibitem[Cho et al.(2016)]{Cho16} Cho, H., Blakeslee, J.~P., Chies-Santos, A.~L., et al.\ 2016, \apj, 822, 95. doi:10.3847/0004-637X/822/2/95

\bibitem[Chung et al.(2020)]{Chung20} Chung, C., Yoon, S.-J., Cho, H., et al.\ 2020, \apjs, 250, 33. doi:10.3847/1538-4365/abb4e6

\bibitem[Chung et al.(2013a)]{Chung13a} Chung, C., Lee, S.-Y., Yoon, S.-J., et al.\ 2013, \apjl, 769, L3. doi:10.1088/2041-8205/769/1/L3

\bibitem[Chung et al.(2013b)]{Chung13b} Chung, C., Yoon, S.-J., Lee, S.-Y., et al.\ 2013, \apjs, 204, 3. doi:10.1088/0067-0049/204/1/3

\bibitem[Chung et al.(2016)]{Chung16} Chung, C., Yoon, S.-J., Lee, S.-Y., et al.\ 2016, \apj, 818, 201. doi:10.3847/0004-637X/818/2/201

\bibitem[Chung et al.(2017)]{Chung17} Chung, C., Yoon, S.-J., \& Lee, Y.-W.\ 2017, \apj, 842, 91. doi:10.3847/1538-4357/aa6f19

\bibitem[Conroy et al.(2009)]{Conroy09} Conroy, C., Gunn, J.~E., \& White, M.\ 2009, \apj, 699, 486. doi:10.1088/0004-637X/699/1/486

\bibitem[de Grijs et al.(2005)]{deGrijs05} de Grijs, R., Anders, P., Lamers, H.~J.~G.~L.~M., et al.\ 2005, \mnras, 359, 874. doi:10.1111/j.1365-2966.2005.08914.x

\bibitem[Du et al.(2021)]{Du21} Du, M., Ho, L.~C., Debattista, V.~P., et al.\ 2021, \apj, 919, 135. doi:10.3847/1538-4357/ac0e98

\bibitem[El-Badry et al.(2019)]{Bardy19} El-Badry, K., Quataert, E., Weisz, D.~R., et al.\ 2019, \mnras, 482, 4528. doi:10.1093/mnras/sty3007

\bibitem[Fahrion et al.(2020)]{Fahrion20} Fahrion, K., Lyubenova, M., Hilker, M., et al.\ 2020, \aap, 637, A27. doi:10.1051/0004-6361/202037686

\bibitem[Forbes \& Remus(2018)]{Forbes18} Forbes, D.~A. \& Remus, R.-S.\ 2018, \mnras, 479, 4760. doi:10.1093/mnras/sty1767

\bibitem[Harris et al.(2004)]{Har04} Harris, G.~L.~H., Geisler, D., Harris, W.~E., et al.\ 2004, \aj, 128, 712. doi:10.1086/421847

\bibitem[Hartigan \& Hartigan(1985)]{Hartigan85} Hartigan, J.~A. \& Hartigan, P.~M. 1985, AnSta, 13, 70. doi:10.1214/aos/1176346577

\bibitem[Hempel et al.(2003)]{Hempel03} Hempel, M., Hilker, M., Kissler-Patig, M., et al.\ 2003, \aap, 405, 487. doi:10.1051/0004-6361:20030598

\bibitem[Jang et al.(2021)]{Jang21} Jang, S., Milone, A.~P., Lagioia, E.~P., et al.\ 2021, \apj, 920, 129. doi:10.3847/1538-4357/ac1861

\bibitem[Jardel et al.(2011)]{Jardel11} Jardel, J.~R., Gebhardt, K., Shen, J., et al.\ 2011, \apj, 739, 21. doi:10.1088/0004-637X/739/1/21

\bibitem[Kang \& Noh(2019)]{Kang19} Kang, Y.-J. \& Noh, Y. 2019, Math. Probl. Eng., 2019, 17. doi:10.1155/2019/4819475

\bibitem[Kim et al.(2002)]{Kim02} Kim, Y.-C., Demarque, P., Yi, S.~K., et al.\ 2002, \apjs, 143, 499. doi:10.1086/343041

\bibitem[Kim et al.(2013)]{HSKim13} Kim, H.-S., Yoon, S.-J., Sohn, S.~T., et al.\ 2013, \apj, 763, 40. doi:10.1088/0004-637X/763/1/40

\bibitem[Kim \& Yoon(2017)]{Kim17} Kim, S. \& Yoon, S.-J.\ 2017, \apj, 843, 43. doi:10.3847/1538-4357/aa7387

\bibitem[Kim et al.(2013)]{Kim13} Kim, S., Yoon, S.-J., Chung, C., et al.\ 2013, \apj, 768, 138. doi:10.1088/0004-637X/768/2/138

\bibitem[Kim et al.(2021)]{Kim21} Kim, S., Yoon, S.-J., Lee, S.-Y., et al.\ 2021, \apjs, 256, 29. doi:10.3847/1538-4365/ac10c2

\bibitem[Knapp(2007)]{Knapp07} Knapp, T. R.\ 2007, JMASM, 6, 8. doi:10.22237/jmasm/1177992120

\bibitem[Kundu \& Zepf(2007)]{Kundu07} Kundu, A. \& Zepf, S.~E.\ 2007, \apjl, 660, L109. doi:10.1086/518214

\bibitem[Landolt(1992)]{Lan92} Landolt, A.~U.\ 1992, \aj, 104, 340. doi:10.1086/116242

\bibitem[Larsen et al.(2001)]{Lars01} Larsen, S.~S., Forbes, D.~A., \& Brodie, J.~P.\ 2001, \mnras, 327, 1116. doi:10.1046/j.1365-8711.2001.04797.x

\bibitem[Lee et al.(2002)]{Lee02} Lee, H.-. chul ., Lee, Y.-W., \& Gibson, B.~K.\ 2002, \aj, 124, 2664. doi:10.1086/344066

\bibitem[Lee et al.(2019)]{Lee19} Lee, S.-Y., Chung, C., \& Yoon, S.-J.\ 2019, \apjs, 240, 2. doi:10.3847/1538-4365/aaecd4

\bibitem[Lee et al.(2020)]{Lee20} Lee, S.-Y., Chung, C., \& Yoon, S.-J.\ 2020, \apj, 905, 124. doi:10.3847/1538-4357/abc4e9

\bibitem[Lee \& Jang(2016)]{Lee16} Lee, M.~G. \& Jang, I.~S.\ 2016, \apj, 819, 77. doi:10.3847/0004-637X/819/1/77

\bibitem[Maechler(2021)]{Maechler21} Maechler, M.\ 2021, diptest: Hartigan's Dip Test Statistic for Unimodality - Corrected. R package version 0.76-0 https://CRAN.R-project.org/package=diptest

\bibitem[Monet et al.(2003)]{Monet03} Monet, D.~G., Levine, S.~E., Canzian, B., et al.\ 2003, \aj, 125, 984. doi:10.1086/345888

\bibitem[Muratov \& Gnedin(2010)]{Muratov10} Muratov, A.~L. \& Gnedin, O.~Y.\ 2010, \apj, 718, 1266. doi:10.1088/0004-637X/718/2/1266

\bibitem[Park \& Lee(2013)]{Park13} Park, H.~S. \& Lee, M.~G.\ 2013, \apjl, 773, L27. doi:10.1088/2041-8205/773/2/L27

\bibitem[Park et al.(2012)]{Park12} Park, H.~S., Lee, M.~G., Hwang, H.~S., et al.\ 2012, \apj, 759, 116. doi:10.1088/0004-637X/759/2/116

\bibitem[Peng et al.(2004)]{Peng04} Peng, E.~W., Ford, H.~C., \& Freeman, K.~C.\ 2004, \apjs, 150, 367. doi:10.1086/381144

\bibitem[Peng et al.(2006)]{Peng06} Peng, E.~W., Jord{\'a}n, A., C{\^o}t{\'e}, P., et al.\ 2006, \apj, 639, 95. doi:10.1086/498210

\bibitem[Peng et al.(2008)]{Peng08} Peng, E.~W., Jord{\'a}n, A., C{\^o}t{\'e}, P., et al.\ 2008, \apj, 681, 197. doi:10.1086/587951

\bibitem[Pfister et al.(2013)]{Pfister13} Pfister, R., Schwarz, K. A., Janczyk, M., et al.\ 2013, Front. Psychol., 4, 700. doi: 10.3389/fpsyg.2013.00700

\bibitem[Pota et al.(2013)]{Pota13} Pota, V., Forbes, D.~A., Romanowsky, A.~J., et al.\ 2013, \mnras, 428, 389. doi:10.1093/mnras/sts029

\bibitem[Puzia et al.(2006)]{Puzia06} Puzia, T.~H., Kissler-Patig, M., \& Goudfrooij, P.\ 2006, \apj, 648, 383. doi:10.1086/505679

\bibitem[Puzia et al.(2005)]{Puzia05} Puzia, T.~H., Kissler-Patig, M., Thomas, D., et al.\ 2005, \aap, 439, 997. doi:10.1051/0004-6361:20047012

\bibitem[Puzia et al.(2002)]{Puzia02} Puzia, T.~H., Zepf, S.~E., Kissler-Patig, M., et al.\ 2002, \aap, 391, 453. doi:10.1051/0004-6361:20020835

\bibitem[R Core Team(2022)]{R22} R Core Team\ 2022, R: A language and environment for statistical computing. R Foundation for Statistical Computing, Vienna, Austria  https://www.R-project.org/.

\bibitem[Raftery(1995)]{Raftery95} Raftery, A.~E.\ 1995, Sociol. Methodol., 25, 111. doi:10.2307/271063.

\bibitem[Rejkuba et al.(2002)]{Rejk02} Rejkuba, M., Minniti, D., Courbin, F., et al.\ 2002, \apj, 564, 688. doi:10.1086/324500

\bibitem[Rejkuba et al.(2011)]{Rejkuba11} Rejkuba, M., Harris, W.~E., Greggio, L., et al.\ 2011, \aap, 526, A123. doi:10.1051/0004-6361/201015640

\bibitem[Richtler et al.(2015)]{Richtler15} Richtler, T., Salinas, R., Lane, R.~R., et al.\ 2015, \aap, 574, A21. doi:10.1051/0004-6361/201424530

\bibitem[Salinas et al.(2015)]{Salinas15} Salinas, R., Alabi, A., Richtler, T., et al.\ 2015, \aap, 577, A59. doi:10.1051/0004-6361/201425574

\bibitem[SAS Institute Inc.(1990)]{SAS90} SAS Institute Inc.\ 1990 SAS/STAT User's Guide, Version 6, 4$^{th}$ Ed. (Cary, NC: SAS Institute), 561

\bibitem[Schlafly \& Finkbeiner(2011)]{Sch11} Schlafly, E.~F. \& Finkbeiner, D.~P.\ 2011, \apj, 737, 103. doi:10.1088/0004-637X/737/2/103

\bibitem[Schlegel et al.(1998)]{Schlegel98} Schlegel, D.~J., Finkbeiner, D.~P., \& Davis, M.\ 1998, \apj, 500, 525. doi:10.1086/305772

\bibitem[Schuberth et al.(2010)]{Schuberth10} Schuberth, Y., Richtler, T., Hilker, M., et al.\ 2010, \aap, 513, A52. doi:10.1051/0004-6361/200912482

\bibitem[Schwarz(1978)]{Schwarz78} Schwarz, G.\ 1978, Ann. Stat., 6, 461. doi:10.1214/aos/1176344136

\bibitem[Scrucca et al.(2016)]{Scrucca16} Scrucca, L., Fop, M., Murphy, T.~B., Raftery, A.~E.\ 2016, 289, mclust 5: clustering, classification and density estimation using Gaussian finite mixture models., The R Journal https://CRAN.R-project.org/package=mclust doi:10.32614/RJ-2016-021

\bibitem[Spitler et al.(2008)]{Spitler08} Spitler, L.~R., Forbes, D.~A., \& Beasley, M.~A.\ 2008, \mnras, 389, 1150. doi:10.1111/j.1365-2966.2008.13681.x

\bibitem[Stetson(1987)]{Stet87} Stetson, P.~B.\ 1987, \pasp, 99, 191. doi:10.1086/131977

\bibitem[Stetson(1990)]{Stet90} Stetson, P.~B.\ 1990, \pasp, 102, 932. doi:10.1086/132719

\bibitem[Stetson(1993)]{Stet93} Stetson, P.~B.\ 1993, IAU Colloq. 136, Stellar Photometry--Current Techniques and Future Developments, ed. C. J. Butler \& I. Elliot (Cambridge: Cambridge Univ. Press), 291

\bibitem[Stetson(1994)]{Stet94} Stetson, P.~B.\ 1994, \pasp, 106, 250. doi:10.1086/133378

\bibitem[Stetson(2000)]{Stet00} Stetson, P.~B.\ 2000, \pasp, 112, 925. doi:10.1086/316595

\bibitem[Strader et al.(2007)]{Strader07} Strader, J., Beasley, M.~A., \& Brodie, J.~P.\ 2007, \aj, 133, 2015. doi:10.1086/512770

\bibitem[Taylor et al.(2017)]{Taylor17} Taylor, M.~A., Puzia, T.~H., Mu{\~n}oz, R.~P., et al.\ 2017, \mnras, 469, 3444. doi:10.1093/mnras/stx1021

\bibitem[Usher et al.(2012)]{Usher12} Usher, C., Forbes, D.~A., Brodie, J.~P., et al.\ 2012, \mnras, 426, 1475. doi:10.1111/j.1365-2966.2012.21801.x

\bibitem[Valdes(1998)]{Valdes98} Valdes, F.~G.\ 1998, in ASP Conf. Ser. 145, Astronomical Data Analysis Software and Systems VII, ed. R. Albrecht, R. N. Hook, \& H. A. Bushouse, 53

\bibitem[VandenBerg et al.(2010)]{Van10} VandenBerg, D.~A., Casagrande, L., \& Stetson, P.~B.\ 2010, \aj, 140, 1020. doi:10.1088/0004-6256/140/4/1020

\bibitem[VandenBerg \& Clem(2003)]{Van03} VandenBerg, D.~A. \& Clem, J.~L.\ 2003, \aj, 126, 778. doi:10.1086/376840

\bibitem[Vanderbeke et al.(2014)]{Vanderbeke14} Vanderbeke, J., West, M.~J., De Propris, R., et al.\ 2014, \mnras, 437, 1734. doi:10.1093/mnras/stt2012

\bibitem[Vazdekis et al.(2015)]{Vazdekis15} Vazdekis, A., Coelho, P., Cassisi, S., et al.\ 2015, \mnras, 449, 1177. doi:10.1093/mnras/stv151

\bibitem[Villaume et al.(2019)]{Vill19} Villaume, A., Romanowsky, A.~J., Brodie, J., et al.\ 2019, \apj, 879, 45. doi:10.3847/1538-4357/ab24d7

\bibitem[Woodley et al.(2007)]{Wood07} Woodley, K.~A., Harris, W.~E., Beasley, M.~A., et al.\ 2007, \aj, 134, 494. doi:10.1086/518788

\bibitem[Worthey(1994)]{Worthey94} Worthey, G.\ 1994, \apjs, 95, 107. doi:10.1086/192096

\bibitem[Worthey \& Lee(2011)]{Worthey11} Worthey, G. \& Lee, H.-. chul .\ 2011, \apjs, 193, 1. doi:10.1088/0067-0049/193/1/1

\bibitem[Yi et al.(2004)]{Yi04} Yi, S.~K., Peng, E., Ford, H., et al.\ 2004, \mnras, 349, 1493. doi:10.1111/j.1365-2966.2004.07614.x

\bibitem[Yoon et al.(2006)]{Yoon06} Yoon, S.-J., Yi, S.~K., \& Lee, Y.-W.\ 2006, Sci, 311, 1129. doi:10.1126/science.1122294

\bibitem[Yoon et al.(2011a)]{Yoon11a} Yoon, S.-J., Lee, S.-Y., Blakeslee, J.~P., et al.\ 2011a, \apj, 743, 150. doi:10.1088/0004-637X/743/2/150

\bibitem[Yoon et al.(2011b)]{Yoon11b} Yoon, S.-J., Sohn, S.~T., Lee, S.-Y., et al.\ 2011b, \apj, 743, 149. doi:10.1088/0004-637X/743/2/149

\bibitem[Yoon et al.(2013)]{Yoon13} Yoon, S.-J., Sohn, S.~T., Kim, H.-S., et al.\ 2013, \apj, 768, 137. doi:10.1088/0004-637X/768/2/137

\end{thebibliography}
\end{document}